\documentclass[seceq,preprint]{ptptex}
\usepackage{graphicx,color,bm}

\newcommand{\C}[1]{\mathcal{#1}}
\newcommand{\T}[1]{\text{#1}}

\allowdisplaybreaks

\begin{document}

%\preprintnumber[\textwidth]{
% Submitted to PTP 2009/1/27
%OU-TAP 293 \hfill PTPTeX ver.0.9}

\markboth{Fujita, Hikida, Tagoshi}{An efficient numerical method for computing gravitational waves...}

\title{
An Efficient Numerical Method for Computing Gravitational Waves 
Induced by a Particle Moving on
Eccentric Inclined Orbits around a Kerr Black Hole
}

\author{Ryuichi Fujita,$^{1,2}$
Wataru Hikida$^2$ and
Hideyuki Tagoshi$^2$}

\inst{
${}^1$Theoretical Physics, Raman Research Institute, Bangalore 560 080, India\\
${}^2$Department of Earth and Space Science, Graduate School of
Science, Osaka University, Toyonaka, 560-0043, Japan\\
}

\recdate{(Received January 27, 2009)}

\abst{
We develop a numerical code to compute
gravitational waves induced by a particle moving on
eccentric inclined orbits around a Kerr black hole.
For such systems, the black hole perturbation method is applicable. 
The gravitational waves can be evaluated by solving the
Teukolsky equation with a point like source term, which is computed from 
the stress-energy tensor of a test particle moving on generic bound geodesic orbits.
In our previous papers, we computed the homogeneous solutions of the
Teukolsky equation using a formalism developed by Mano, Suzuki and
Takasugi and showed that we could 
compute gravitational waves efficiently and very accurately 
in the case of circular orbits on the equatorial plane.
Here, we apply this method to eccentric inclined orbits.
The geodesics around a Kerr black hole have three constants of motion:
energy, angular momentum and the Carter constant.
We compute the rates of change of the Carter 
constant as well as those of energy and angular momentum. 
This is the first time that the rate of change of the Carter constant has been 
evaluated accurately. 
We also treat the case of highly eccentric orbits with $e=0.9$. 
To confirm the accuracy of our codes, several tests are performed. 
We find that the accuracy is only limited by the truncation of 
$\ell$-, $k$- and $n$-modes, where $\ell$ is the index of the spin-weighted 
spheroidal harmonics, and $n$ and $k$ are the harmonics of the radial and polar motion,
respectively. 
When we set the maximum of $\ell$ to $20$, we obtain 
a relative accuracy of $10^{-5}$ even in the highly eccentric case of $e=0.9$.
The accuracy is better for lower eccentricity.
Our numerical code is expected to be useful for computing 
templates of the extreme mass ratio inspirals, which is one of the 
main targets of the Laser Interferometer Space Antenna (LISA). 
}

\maketitle

%%%%%%%%%%%%%%%%%%%%%%%%%%%%%%%%%%%%%%%%%%%%%%%%%%%%%%%%%%%%%%%%
\section{Introduction}\label{sec:introduction}
%%%%%%%%%%%%%%%%%%%%%%%%%%%%%%%%%%%%%%%%%%%%%%%%%%%%%%%%%%%%%%%%

Gravitational radiation is one of the most important predictions resulting from
general relativity. 
The presence of gravitational radiation 
has been proved indirectly through its effect
on the orbital period of the Hulse-Taylor binary pulsar~\cite{Hulse:1975}.
Owing to advances in modern technology, 
the era of gravitational wave astronomy has almost arrived. 
Several ground-based interferometric gravitational wave detectors have been 
operated~\cite{LIGO,TAMA,GEO,VIRGO} for several years. 
Next-generation ground-based interferometers
such as advanced LIGO, advanced VIRGO and LCGT, 
are planned, and will be started in the near future.  
Research and development studies of a space-based gravitational wave observatory project,
the Laser Interferometer Space Antenna (LISA)~\cite{LISA}, are 
rapidly progressing. There are also proposals for 
laser interferometer gravitational wave antennas in space
such like a DECihertz Interferometer Gravitational wave Observatory
(DECIGO)~\cite{DECIGO} and a Big Bang Observer (BBO)~\cite{BBO}. 
Those will be sensitive to frequencies of $10^{-2}\leq f \leq 1\ {\rm Hz}$.

One of the most promising sources of gravitational waves 
that can be detected by LISA
is a compact star orbiting a supermassive black hole. 
Observing gravitational waves from this type of binary system, i.e., an extreme
mass ratio inspiral (EMRI), we may be able to obtain information 
on the central black hole's spacetime such as the mass, spin of the black hole
and the mass distribution of compact objects in the center of the galaxy. 
To extract the physical information of EMRI from 
observational data obtained by the detectors, 
we have to compute the theoretical waveforms with a 
phase accuracy within one cycle over the total number of cycles.
LISA is sensitive to gravitational waves around $10^{-2}$ Hz. 
When the observation of LISA is performed for one year, 
the total number of cycles of waves is $\sim 10^{5}$. 
Thus, to analyze one year of data obtained from LISA, we need theoretical waveforms
that are accurate to $10^{-5}$. 

The dynamics of EMRI is accurately modeled as a point particle of small
mass moving around a Kerr black hole.
Therefore, gravitational waves from EMRI can be evaluated using black
hole perturbation theory, 
which was originally developed as a metric
perturbation theory for a black hole spacetime. 
For nonrotating (Schwarzschild) black holes, a single master equation
for the metric perturbation was derived by
Regge and Wheeler for the odd-parity parts~\cite{Regge:1957td}, 
and later by Zerilli for the even-parity parts~\cite{Zerilli:1971wd}. 
These equations are remarkably simple, separable,
hyperbolic equations with a potential term. 
For rotating black holes, at present, there
are no simple decoupled equations for metric
coefficients. 
Instead the perturbed geometry must be analyzed by the
equations derived by Teukolsky using gauge-invariant 
variables corresponding to some tetrad components of the perturbed Weyl
curvature~\cite{Teukolsky:1973ha}.

Using this Teukolsky formalism, there have been many numerical computations
of gravitational waves induced by a point particle. See
Chandrasekhar~\cite{Chandra} and Nakamura {\it et al.}~\cite{Nakamura:1987}, 
for reviews and for references on earlier papers.
For simple orbits such as circular or equatorial orbits 
around a black hole, an accuracy of $10^{-5}$ has been achieved, which may be 
sufficient to detect gravitational waves. 
Computations of gravitational waves induced by a particle moving
on eccentric nonequatorial orbits are now available~\cite{Hughes:2005qb}.
However, a large amount of computation time is needed and the results have only
a few orders of accuracy.

Furthermore, it has been pointed out that 
EMRIs radiating gravitational waves within the LISA band
have high eccentricities~\cite{Hopman:2005vr}.
So far, the gravitational energy flux has been computed only for 
orbits with eccentricity less than $0.7$.
This is primarily due to the low numerical accuracy and high computational time
at the high-frequency modes. 
In the Teukolsky formalism in the frequency domain,
when we treat a large eccentricity and large inclination angle,  
we need to compute a large number of harmonics corresponding to the radial 
and polar motion. Thus, the computation time becomes a serious problem.
It is thus desirable to develop more efficient and more accurate numerical codes.

In the Kerr spacetime, there are three constants of motion of a geodesic:
the energy, angular momentum and Carter constant. 
Using the conservation law, we can evaluate the rates of change of
the energy and angular momentum of a particle due to the emission of gravitational waves.
In contrast, the rate of change of the Carter constant cannot be derived from the conservation law.
Mino proposed a method to evaluate the average rates of change 
of the three constants including the Carter constant~\cite{Mino:2003yg} under an adiabatic approximation. 
He showed that the average rates of change can be evaluated 
using the radiative field instead of the retarded field. 
Using Mino's method, Drasco {\itshape et al.}~\cite{DFH} derived 
simplified version of his formula for the case of a scalar field. 
Sago {\itshape et al.}~\cite{Sago:2005gd} developed Mino's method further, giving a
simplified version of his formula.
Applying their new scheme, they gave explicit analytic formulas for
the rates of change of constants when a particle moves on 
slightly eccentric and slightly inclined orbits~\cite{Sago:2005fn}.
Ganz {\itshape et al.} extended this computation to the case when 
a particle moves on slightly eccentric and arbitrarily inclined orbits~\cite{Ganz}. 
However, these two results are based on the assumption of small eccentricity
and the post-Newtonian approximation to the order $O((v/c)^5)$.

In this paper, we compute the rates of change 
of the three constants of motion, including the Carter constant, induced by
a particle moving on eccentric and inclined orbits around a Kerr black
hole without assuming small eccentricities, small inclination angles or low velocity. 
This is the first time that the rate of change of the
Carter constant has been computed accurately using an adiabatic approximation.
Drasco and Hughes~\cite{Drasco:2006} computed the rate of change 
of the Carter constant assuming that the inclination angle does not change.
However, this assumption holds only approximately, rather than exactly. 

To treat cases of large eccentricity and inclination angle, 
we introduce various methods. 
The numerical method used to compute the homogeneous solution 
of the Teukolsky equation is based on that of Fujita and 
Tagoshi~\cite{Fujita:2004rb,Fujita:2005rb}, in which the Mano-Suzuki-Takasugi
formalism \cite{Mano:1996vt} is used. However, since only circular, equatorial orbits are 
treated in these works,
we extend the method so that we can compute the homogeneous solutions 
efficiently in eccentric and inclined cases.
For the source term, 
we introduce analytical expressions for the radial and polar motion 
to achieve high accuracy.

The paper is organized as follows.
In \S~\ref{sec:grav-waves-from}, we summarize the details of the method
for computing the gravitational waves from EMRIs using black hole
perturbation theory.
In \S~\ref{sec:our-numerical-method}, we explain our numerical methods.
In \S~\ref{sec:results}, we 
discuss the peaks of radial and polar modes briefly then
we derive the rates of change of the three constants of motion for
highly eccentric orbits.
Next, we verify our code using the Schwarzschild case and by comparison with analytical
post-Newtonian expressions. 
Section~\ref{sec:summary} is devoted to a summary and
discussion.
In the Appendices, some detailed formulas are given. 
Throughout this paper we use units with $c=G=1$.

%%%%%%%%%%%%%%%%%%%%%%%%%%%%%%%%%%%%%%%%%%%%%%%%%%%%%%%%%%%%%%%%
\section{Formulas for the rate of change due to gravitational wave emission}
\label{sec:grav-waves-from}
%%%%%%%%%%%%%%%%%%%%%%%%%%%%%%%%%%%%%%%%%%%%%%%%%%%%%%%%%%%%%%%%
In the Teukolsky formalism, 
the gravitational perturbation of a Kerr black hole is described in terms of 
the Newman-Penrose variables, $\Psi_0$ and $\Psi_4$, which satisfy  the master equation.
The Weyl scalar $\Psi_4$
is related to the amplitude of the gravitational wave at infinity as
\begin{eqnarray}
\Psi_4\rightarrow\frac{1}{2}(\ddot{h}_{+}-i\,\ddot{h}_{\times }),\,\,\,{\rm for}\,\,\,r\rightarrow\infty.
\end{eqnarray}
The master equation for $\Psi_4$
can be separated into radial and angular parts if we
expand $\Psi_4$ in Fourier harmonic modes as
\begin{eqnarray}
\rho^{-4} \Psi_4=\displaystyle \sum_{\ell m}\int_{-\infty}^{\infty} d\omega 
e^{-i\omega t+i m \varphi} \ _{-2}S_{\ell m}^{a\omega}(\theta)
R_{\ell m\omega}(r),
\end{eqnarray}
where $\rho=(r-i a \cos\theta)^{-1}$, 
the angular function $_{-2}S_{\ell m}^{a\omega}(\theta)$ 
is the spin-weighted spheroidal harmonic with spin $s=-2$,  and $M$ and $aM$ are
the mass and angular momentum of the black hole, respectively. 
The radial function $R_{\ell m\omega}(r)$ satisfies 
the radial Teukolsky equation, 
\begin{eqnarray}
\Delta^2\frac{d}{dr}\left(\frac{1}{\Delta}\frac{dR_{\ell m\omega}}{dr}
\right)-V(r) R_{\ell m\omega}=T_{\ell m\omega},
\label{eq:radial-teukolsky}
\end{eqnarray}
where $\Delta=r^2-2Mr+a^2$. 
The potential term $V(r)$ is given as
\begin{eqnarray}
V(r) = -\frac{K^2 - 2is(r-M)K}{\Delta} - 4is\omega r + \lambda,
\end{eqnarray}
where $K=(r^2+a^2)\omega-ma$ and $\lambda$ is the 
eigenvalue of $_{-2}S_{\ell m}^{a\omega}(\theta)$.

We solve the radial Teukolsky equation by using 
the Green function method. The solution of the Teukolsky equation 
with a purely outgoing property at infinity and a purely ingoing 
property at the horizon becomes
\begin{align}
R_{\ell m \omega}(r)=
\frac{1}{W_{\ell m \omega}}&\left\{
R^{\rm up}_{\ell m\omega}(r)
\int_{r_+}^r dr' R^{\rm in}_{\ell m \omega}T_{\ell m \omega}\Delta^{-2}\right.\cr
&\left.+R^{\rm in}_{\ell m\omega}(r)
\int_{r}^\infty dr' R^{\rm up}_{\ell m \omega}T_{\ell m \omega}\Delta^{-2}
\right\},
\label{Rfield}
\end{align}
where the Wronskian $W_{\ell m \omega}$ is given as
\begin{eqnarray}
W_{\ell m \omega}=2i\omega C^{\rm trans}_{\ell m \omega} B^{\rm inc}_{\ell m \omega},
\end{eqnarray}
and where $R^{\rm in/up}_{\ell m\omega}(r)$
satisfy ingoing/outgoing wave conditions at the horizon/infinity.
The asymptotic forms of $R^{\rm in/up}_{\ell m\omega}(r)$ are expressed as
\begin{align}
& R^{\T{in}}_{\ell m\omega} \rightarrow \left\{
\begin{array}{ll}
 B^{\T{trans}}_{\ell m\omega}\Delta^{2}e^{-iP r^{*}}& \T{for } r\to r_+,\\
 r^{3}B^{\T{ref}}_{\ell m\omega}e^{i\omega r^{*}}
+r^{-1}B^{\T{inc}}_{\ell m\omega}e^{-i\omega r^{*}}
&\T{for }r\to\infty, \\
\end{array}
\right. \cr
& R^{\T{up}}_{\ell m\omega} \rightarrow \left\{
\begin{array}{ll}
 C^{\T{up}}_{\ell m\omega}e^{i P r^{*}}
+\Delta^{2}C^{\T{ref}}_{\ell m\omega}e^{-i P r^{*}}
& \T{for } r\to r_+,\\
r^3 C^{\T{trans}}_{\ell m\omega} e^{ i\omega r^{*}}
&\T{for }r\to\infty,\\
\end{array}
\right. 
\end{align}
where $P=\omega - ma/2Mr_+$ and $r^*$ is the tortoise coordinate defined as
\begin{align}
 r^* = r + \frac{2Mr_+}{r_+-r_-}\ln\frac{r-r_+}{2M}
- \frac{2Mr_-}{r_+-r_-}\ln\frac{r-r_-}{2M},
\end{align}
with $r_{\pm}=M\pm\sqrt{M^2-a^2}$. 

The asymptotic property of the solution at the horizon is expressed as
\begin{eqnarray}
R_{\ell m \omega}(r\rightarrow r_+)=
\frac{B^{\rm trans}_{\ell m \omega}\Delta^2e^{-iPr^*}}
{2i\omega C^{\rm trans}_{\ell m \omega}B^{\rm inc}_{\ell m \omega}}
\int_{r_+}^\infty dr' R^{\rm up}_{\ell m \omega}T_{\ell m \omega}\Delta^{-2}
\equiv Z^{\rm H}_{\ell m \omega}\Delta^2e^{-iPr^*}.
\end{eqnarray}
The solution at infinity is expressed as 
\begin{eqnarray}
R_{\ell m \omega}(r\rightarrow \infty)=
\frac{r^3e^{i\omega r^*}}{2i\omega B^{\rm inc}_{\ell m \omega}}
\int_{r_+}^{\infty}dr'
R^{\rm in}_{\ell m \omega}T_{\ell m \omega}\Delta^{-2}
\equiv Z^\infty_{\ell m \omega}r^3e^{i\omega r^*}.
\end{eqnarray}

Using the formula of the source term $T_{\ell m \omega}$~\cite{chapter}, 
$Z_{\ell m\omega}^{\infty,{\rm H}}$ are expressed as
\begin{align}
Z^{\T{H}}_{\ell m\omega} &= \frac{\mu B^{\T{trans}}_{\ell m\omega}}
{2i\omega C^{\T{trans}}_{\ell m\omega}B^{\T{inc}}_{\ell m\omega}}
\int^{\infty}_{-\infty}dt e^{i\omega t-im\phi(t)} \C{I}_{\ell
m\omega}^{\T{H}}[r(t),\theta(t)],\cr
Z^{\infty}_{\ell m\omega} &= \frac{\mu}{2i\omega B^{\T{inc}}_{\ell m\omega}}
\int^{\infty}_{-\infty}dt e^{i\omega t-im\phi(t)} \C{I}_{\ell
m\omega}^{\infty}[r(t),\theta(t)],
\label{eq:Z1}
\end{align}
where 
\begin{align}
 \C{I}^{\T{H}}_{\ell m\omega} =&
\left[
R^{\T{up}}_{\ell m\omega}\left\{
A_{nn0}+A_{\bar{m}n0}+A_{\bar{m}\bar{m}0}
\right\}\right.\cr
&\left.-
\frac{dR^{\T{up}}_{\ell m\omega}}{dr}
\left\{
A_{\bar{m}n1}+A_{\bar{m}\bar{m}1}
\right\}
+
\frac{d^2R^{\T{up}}_{\ell m\omega}}{d^2r}
A_{\bar{m}\bar{m}2}
\right]_{r=r(t),\theta=\theta(t)},\cr
 \C{I}^{\infty}_{\ell m\omega} =&
\left[
R^{\T{in}}_{\ell m\omega}\left\{
A_{nn0}+A_{\bar{m}n0}+A_{\bar{m}\bar{m}0}
\right\}\right.\cr
&\left.-
\frac{dR^{\T{in}}_{\ell m\omega}}{dr}
\left\{
A_{\bar{m}n1}+A_{\bar{m}\bar{m}1}
\right\}
+
\frac{d^2R^{\T{in}}_{\ell m\omega}}{d^2r}
A_{\bar{m}\bar{m}2}
\right]_{r=r(t),\theta=\theta(t)},
\end{align}
where $A_{nn0}$ and other terms are given in Appendix~\ref{sec:teukolsky-formalism}.

The function $\C{I}_{\ell m\omega}^{\infty,{\rm H}}[r(t),\theta(t)]$ 
is constructed from the source 
term of the Teukolsky equation and depends on the orbital
worldline of the star perturbing the black hole spacetime. 
If the trajectory of a compact star is eccentric and inclined from the equatorial 
plane, it is not easy  to evaluate Eq. (\ref{eq:Z1}).  
This is because the radial and polar motion are coupled in the observer time variable $t$. 
This problem can be solved by introducing 
a new time variable $\lambda$ defined as $d\lambda=d\tau/\Sigma$,
where $\Sigma=r^2+a^2\cos^2 \theta$ \cite{Mino:2003yg,Drasco:2004}. 
The geodesic equations become
\begin{eqnarray}
\left(\frac{dr}{d\lambda}\right)^2
&=&\left[(r^2+a^2)\C{E}-a \C{L}_{z}\right]^2
             - \Delta[r^2+(\C{L}_{z}-a \C{E})^2+\C{C}]
\equiv R(r),\label{geodesic_r}
\\
\left(\frac{d\cos\theta}{d\lambda}\right)^2 &=&
\C{C} - (\C{C}+a^2(1-\C{E}^2)+\C{L}_{z}^2)\cos^2\theta
+ a^2(1-\C{E}^2)\cos^4\theta\nonumber
\\
&\equiv&\Theta(\cos\theta),
\label{geodesic_theta}
\\
\frac{d\phi}{d\lambda} &=&
\Phi_{\rm r}(r)+\Phi_\theta(\cos\theta) - a\C{E},
\quad \quad 
\label{geodesic_phi}
\\
\frac{dt}{d\lambda} &=& 
T_r(r)+T_\theta(\cos\theta) + a\C{L}_{z},
\label{geodesic_t}
\end{eqnarray}
where
\begin{eqnarray*}
&&
\Phi_{\rm r}(r)\equiv \frac{a}{\Delta}\left[\C{E}(r^{2}
+a^{2}) - a \C{L}_{z}\right], \quad 
\Phi_\theta(\cos\theta) \equiv \frac{\C{L}_{z}}{1-\cos^{2}\theta}, \\
&&T_r(r)\equiv \frac{r^{2}+a^{2}}{\Delta}
\left[\C{E}(r^{2}+a^{2}) - a \C{L}_{z}\right], \quad
T_\theta(\cos\theta)\equiv - a^2 \C{E}(1-\cos^{2}\theta),
\end{eqnarray*}
and $\C{E}$, $\C{L}_{z}$ and $\C{C}$ 
are the energy, the $z$-component of the angular 
momentum and the Carter constant per unit mass, respectively. 
The equations of radial and polar motion are decoupled in Eqs. (\ref{geodesic_r})
and (\ref{geodesic_theta}).
For the bound orbits, $r(\lambda)$ and $\theta(\lambda)$ become
periodic functions that are independent of each other. 
The fundamental periods for the radial and polar motion,
$\Lambda_r$ and $\Lambda_\theta$, are defined as
\begin{align}
\Lambda_r=2\int_{\rm r_{\rm min}}^{\rm r_{\rm max}}\frac{dr}{\sqrt{R(r)}},\quad
\Lambda_\theta=4\int_{0}^{\cos\theta_{\rm min}}
\frac{d\cos\theta}{\sqrt{\Theta(\cos\theta)}}.
\label{periods}
\end{align}
The angular frequencies of the radial and polar motion become
\begin{equation}
\Upsilon_r=\frac{2\pi}{\Lambda_r},\quad\quad
\Upsilon_\theta=\frac{2\pi}{\Lambda_\theta}.
\label{frequencies}
\end{equation}
Explicit expressions for $\Upsilon_r$ and $\Upsilon_\theta$ are given in
Appendix \ref{sec:geodesic-motion-kerr}.

We define the angle variables as $w_r=\Upsilon_r\lambda$ and $w_\theta=\Upsilon_\theta\lambda$. 
The functions that depend only on $r$ or $\theta$ become
periodic functions with respect to $w_r$ or $w_\theta$, respectively,
with period $2\pi$. 

%%%%%%%%%%%%%%%%%%%%%%%%%%%%%%%%%%%%%%%%%%%%%%%%%%%%%%%%%%%%%%%%%%%%%%%%%%
We expand the right-hand sides of Eqs.~(\ref{geodesic_phi}) and (\ref{geodesic_t})
into Fourier series,
\begin{eqnarray}
\label{geodesic_phi_fourier}
\frac{dt}{d\lambda} &=&\sum_{k,n}T_{k,n}
e^{-ik\Upsilon_r\lambda}e^{-in\Upsilon_\theta\lambda}, \\
\label{geodesic_t_fourier}
\frac{d\phi}{d\lambda} &=&\sum_{k,n}\Phi_{k,n}
e^{-ik\Upsilon_r\lambda}e^{-in\Upsilon_\theta\lambda},
\end{eqnarray}
where
\begin{eqnarray}
T_{k,n}&=&\frac{1}{(2\pi)^2}\int_0^{2\pi}dw_r \int_0^{2\pi}dw_\theta
(T_r(r)+T_\theta(\cos\theta)+a\C{L}_z)
e^{i k w_r}e^{i n w_\theta}, \\
\Phi_{k,n}&=&\frac{1}{(2\pi)^2}\int_0^{2\pi}dw_r \int_0^{2\pi}dw_\theta
(\Phi_r(r)+\Phi_\theta(\cos\theta)-a\C{E})
e^{i k w_r}e^{i n w_\theta}. \\
\end{eqnarray}
Since $T_{k,n}=0$ and $\Phi_{k,n}=0$ in the case of $k\neq 0$ and $n\neq 0$,
we have
\begin{eqnarray}
\frac{dt}{d\lambda} &=&
\Gamma+\sum_{k\neq 0}T_{k,0}e^{-ikw_r}+\sum_{n\neq 0}T_{0,n}e^{-inw_\theta}, \\
\Gamma&\equiv& T_{00} \nonumber\\
&=&\Upsilon_{t^{(r)}}+\Upsilon_{t^{(\theta)}}+a \C{L}_z,\\
\frac{d\phi}{d\lambda} &=&
\Upsilon_\phi+\sum_{k\neq 0}\Phi_{k,0}e^{-ikw_r}+\sum_{n\neq 0}\Phi_{0,n}e^{-inw_\theta}, \\
\Upsilon_\phi&\equiv& \Phi_{00} \nonumber\\
&=&\Upsilon_{\phi^{(r)}}+\Upsilon_{\phi^{(\theta)}}-a \C{E},
\end{eqnarray}
where
\begin{align}
\Upsilon_{t^{(r)}}&=\frac{1}{2\pi}\int_0^{2\pi}dw_r T_r,\,\,\,
\Upsilon_{t^{(\theta)}}=\frac{1}{2\pi}\int_0^{2\pi}dw_\theta T_\theta,\cr
\Upsilon_{\phi^{(r)}}&=\frac{1}{2\pi}\int_0^{2\pi}dw_r \Phi_r,\,\,\,
\Upsilon_{\phi^{(\theta)}}=\frac{1}{2\pi}\int_0^{2\pi}dw_\theta \Phi_\theta.
\end{align}

Since $T_{k,n}=0$ for $k\neq 0$ and $n\neq 0$, 
we need to compute $T_{k,0}, T_{0,n}, \Phi_{k,0}$ and $\Phi_{0,n}$,
which can be easily obtained from a one-dimensional Fourier transformation. 
We obtain the functions $t(\lambda)$ and $\phi(\lambda)$
from the following formulas:
\begin{eqnarray}
t(\lambda)&=&\Gamma\lambda
+\sum_{k\neq 0}\frac{iT_{k,0}}{k\Upsilon_r}e^{-ikw_r}
+\sum_{n\neq 0}\frac{iT_{0,n}}{n\Upsilon_\theta}e^{-inw_\theta}, 
\label{geodesic_t_fourier2} \\
\phi(\lambda)&=&\Upsilon_\phi\lambda
+\sum_{k\neq 0}\frac{i\Phi_{k,0}}{k\Upsilon_r}e^{-ikw_r}
+\sum_{n\neq 0}\frac{i\Phi_{0,n}}{n\Upsilon_\theta}e^{-inw_\theta}.
\label{geodesic_phi_fourier2} 
\end{eqnarray}
The two variables, $\Gamma$ and $\Upsilon_\phi$, represent 
the average rates of change of $t$ and $\phi$ as functions of $\lambda$, 
respectively. 

When we consider the bound orbits of a point particle using $\lambda$, 
the amplitude of the partial wave
$Z^{\infty/\T{H}}_{\ell m\omega}$, defined in
Eq.~\eqref{eq:Z1}, can be expanded by using a Fourier series as
\begin{eqnarray}
Z^{\infty,{\rm H}}_{\ell m\omega}
\equiv \sum_{kn} \tilde{Z}^{\infty,{\rm H}}_{\ell mkn}
\delta(\omega - \omega_{mkn})\;,
\label{zdelta}
\end{eqnarray}
where 
\begin{eqnarray}
\label{eq:zlmkn}
\tilde{Z}^{\infty,{\rm H}}_{\ell mkn} = \frac{1}{(2\pi)^2}\int_0^{2\pi}dw_\theta \int_0^{2\pi}dw_r~e^{i(k w_\theta + n w_r)}\,
Z^{\infty,{\rm H}}_{\ell m\omega_{mkn}}[r(w_r),\theta(w_\theta)]
\end{eqnarray}
and
\begin{equation} \label{omega_mkn}
\omega_{mkn}\equiv (m\Upsilon_\phi + k\Upsilon_\theta + n\Upsilon_r)/\Gamma\;.
\end{equation}

Using these functions, the gravitational waveform at infinity is expressed as
\begin{align}
 h_{+}-ih_{\times}=-\frac{2}{r}\sum_{\ell mkn}
\frac{\tilde{Z}^{\infty}_{\ell mkn}}{\omega_{mkn}^2}
\frac{{}_{-2}S^{a\omega_{mkn}}_{\ell m}(\theta)}{\sqrt{2\pi}}
e^{i\omega_{mkn}(r^*-t)+im\phi}.
\end{align}
Moreover, the time-averaged rates of change for the three constants of
motion due to the emission of gravitational waves are
expressed as~\cite{Mino:2003yg,Sago:2005gd,Sago:2005fn}
\begin{eqnarray}
 \left<\frac{d\C{E}}{dt}\right>
&=& -\mu^{-1}\sum_{\ell mkn}
\frac{1}{4\pi\omega^2_{mkn}}\left(
\left|\tilde{Z}^{\infty}_{\ell mkn}\right|^2
+\alpha_{\ell mkn}
\left|\tilde{Z}^{\T{H}}_{\ell mkn}\right|^2
\right), \label{eq:dEdt} \\
 \left<\frac{d\C{L}_{z}}{dt}\right> &=& -\mu^{-1}\sum_{\ell mkn}
\frac{m}{4\pi\omega^3_{mkn}}
\left(
\left|\tilde{Z}^{\infty}_{\ell mkn}\right|^2
+\alpha_{\ell mkn}
\left|\tilde{Z}^{\T{H}}_{\ell mkn}\right|^2
\right),\label{eq:dLzdt} \\
\left<\frac{d\C{C}}{dt}\right>
&=& \left<\frac{d\C{Q}}{dt}\right>-2(a\C{E}-\C{L}_{z})\left(
a\left<\frac{d\C{E}}{dt}\right>-\left<\frac{d\C{L}_{z}}{dt}\right>
\right), \label{eq:dCdt} \\
\left<\frac{d\C{Q}}{dt}\right> &=&
2\Upsilon_{t^{(r)}}
\left<\frac{d\C{E}}{dt}\right>
-2\Upsilon_{\phi^{(r)}}
\left<\frac{d\C{L}_{z}}{dt}\right>\nonumber\\
&&+\mu^{-2}\sum_{\ell mkn}\frac{n\Upsilon_r}
{2\pi\omega^3_{mkn}}\left(
\left|\tilde{Z}^{\infty}_{\ell mkn}\right|^2
+\alpha_{\ell mkn}
\left|\tilde{Z}^{\T{H}}_{\ell mkn}\right|^2
\right),
\label{eq:dIdt}
\end{eqnarray}
where 
\begin{align}
 \alpha_{\ell mkn} =
 \frac{256(2Mr_+)^5 P (P^2+4\epsilon^2)(P^2+16\epsilon^2)\omega_{mkn}^3}
{C^{\T{TS}}_{\ell mkn}},
\label{eq:TS_const}
\end{align}
where $\epsilon=\sqrt{M^2-a^2}/4Mr_{+}$, 
$P=\omega_{mkn}-ma/(2Mr_+)$, $r_+=M+\sqrt{M^2-a^2}$ is the location of the
event horizon and $C^{\T{TS}}_{\ell mkn}$ is the 
Teukolsky-Starobinsky constant\cite{TPress}.
Here $\left<\cdots\right>$ represents the time average. 

\begin{figure}[tbp]
\begin{center}
\includegraphics[scale=1]{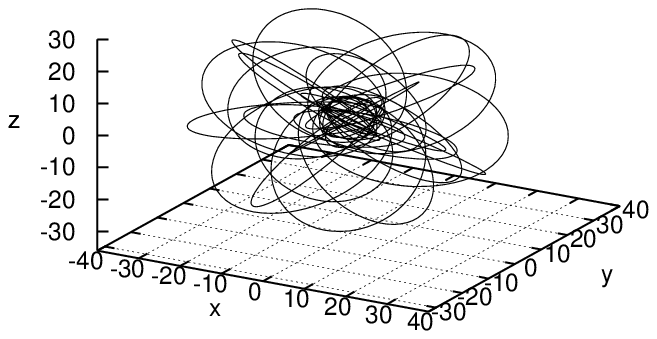}$\quad$
\includegraphics[scale=1]{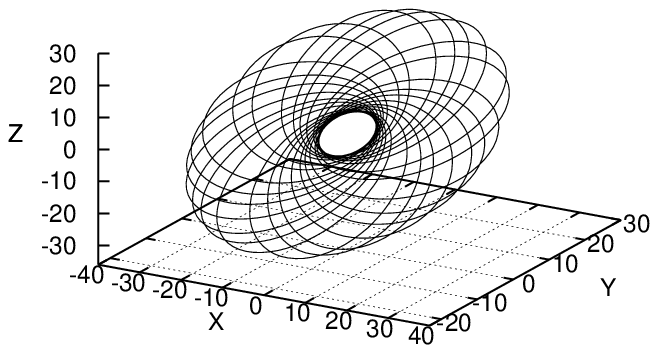}
\end{center}
\caption{Plots of generic bound geodesic orbits with
eccentricity $e=0.7$, semilatus rectum $p=10M$ and inclination angle
$\theta_{\T{inc}}=45^{\circ}$.
The black hole's spin is set to $a=0.9M$.
The left figure is expressed using a Cartesian coordinate system.
The right figure is expressed by using a corotational system, defined as
Eq.~\eqref{eq:co-rotational}.
}
\label{fig:orbit00}
\begin{center}
\includegraphics[scale=1.2]{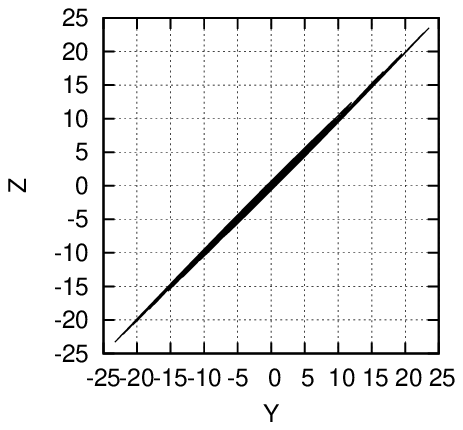}$\qquad$
\includegraphics[scale=1.2]{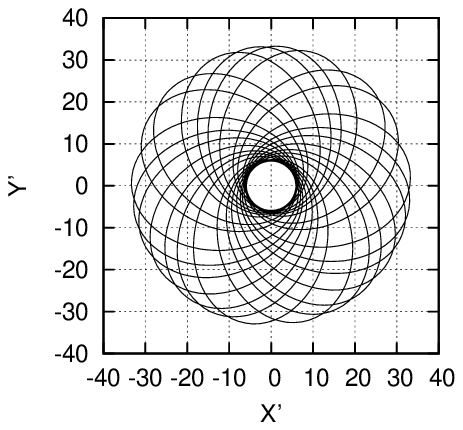}
\end{center}
\caption{The left figure is the projected image of right figure in Fig.~\ref{fig:orbit00}.
The right figure is the projected image of the same figure 
to the $Z=Y\tan\theta_{\T{inc}}$ plane. 
Here the coordinates $(X',Y')$ are defined as $X'=X$ and 
$Y'=Y\cos\theta_{\rm inc}$ $+Z\sin\theta_{\rm inc}$.
The left figure shows that the orbit almost stays on the plane
inclined by $\theta_{\T{inc}}$ from the equatorial plane.
}
\label{fig:orbit01}
\end{figure}

%%%%%%%%%%%%%%%%%%%%%%%%%%%%%%%%%%%%%%%%%%%%%%%%%%%%%%%%%%%%%%%%
\section{Numerical methods}\label{sec:our-numerical-method}
%%%%%%%%%%%%%%%%%%%%%%%%%%%%%%%%%%%%%%%%%%%%%%%%%%%%%%%%%%%%%%%%
In this section, we discuss the numerical method used to compute
$\tilde{Z}^{\infty/\T{H}}_{\ell mkn}$, Eq.~\eqref{eq:zlmkn}.
The numerical computation is divided into four parts: geodesics, homogeneous solutions, 
the integration of Eq.~\eqref{eq:zlmkn} and the mode summation 
in Eqs. (\ref{eq:dEdt})--(\ref{eq:dIdt}).
In the following subsections, we explain the method for obtaining the geodesic and 
homogeneous solutions. 
The numerical integration of Eq.~\eqref{eq:zlmkn}
and the mode summation are explained in \S~\ref{sec:results}. 

%%%%%%%%%%%%%%%%%%%%%
\subsection{Geodesics}\label{sec:geodesic}
%%%%%%%%%%%%%%%%%%%%%
Geodesics around a Kerr black hole are completely specified by the three constants of motion,
$(\C{E},\C{L}_{z},\C{C})$.
Instead of using these three constants, 
it is convenient to introduce three orbital parameters: 
eccentricity $e$, semilatus rectum $p$ and inclination angle
$\theta_{\T{inc}}$
which are defined in Eq.~(\ref{eq:epthetainc}).
There is a one-to-one correspondence between the orbital parameters
$(p,{e},\theta_{\T{inc}})$ and $(\C{E},\C{L}_{z},\C{C})$
\cite{Schmidt:2002, Drasco:2006}.
 
Our method of computing geodesic motion is summarized as follows.
We specify three orbital parameters, $(p,{e},\theta_{\T{inc}})$. 
We first compute the constants of motion, $\C{E},$ $\C{L}_{z}$ and $\C{C}$, 
which correspond to $(p,{e},\theta_{\T{inc}})$. 
We then compute the orbital frequencies of the radial and polar motion, 
$\Upsilon_{r}$ and $\Upsilon_{\theta}$.
We find that the solutions of the $r$ and $\theta$ components of the 
geodesic equations are expressed in terms of Jacobi elliptic functions,
which can be easily computed numerically (see Appendix~\ref{sec:geodesic-motion-kerr}). 
This enables us to evaluate the bound orbits accurately.
We have not confirmed whether this method is faster than the numerical 
integration methods when the same accuracy is required. 
However, since we need to compute the geodesic motion only once, its computation time
is negligible compared with the total computation time. 
Thus, we adopt this method. 
The $r$ and $\theta$ components of coordinates and velocity of the particle, 
$r(w_r)$, $[{dr}/{d\lambda}](w_r)$,$[\cos\theta](w_\theta)$
and $[d\cos\theta/d\lambda](w_\theta)$, 
are respectively expressed using $w_r$ and $w_\theta$ analytically.

The $t$ and $\phi$ components of the solution of the geodesic equations
can be obtained using the Fourier series expansion,
Eqs.~(\ref{geodesic_t_fourier2}) and (\ref{geodesic_phi_fourier2}).
We truncate the Fourier series expansion at a certain frequency.
Since the convergence of the Fourier series is rapid,
as mentioned in Ref.~\citen{Drasco:2006}, this does not cause serious problems.
Explicit expressions for orbital frequencies and 
the solutions of radial and polar motion
are given in Appendix~\ref{sec:geodesic-motion-kerr}.

Here we demonstrate a geodesic orbit computed using our code.
We start an orbit at
$(t,\,r,\,\theta,\,\phi)=(0,\,p/(1+{e}),\,0,\,0)$, and end it after
several oscillation periods of radial and polar motion.
In Fig.~\ref{fig:orbit00}, we plot the orbit on two different
coordinate systems.
The left figure of Fig.~\ref{fig:orbit00} is expressed using
the Cartesian coordinate system $(t,x,y,z)$ defined as
\begin{align}
t=t,\quad
x=r\sin\theta\cos\phi,\quad
y=r\sin\theta\sin\phi,\quad
z=r\cos\theta.
\end{align}
This figure indicates the well-known fact that 
generic geodesic orbits around a Kerr black
hole are complicated.
To understand the physical meaning of the orbital
parameters $p$, $e$ and $\theta_{\T{inc}}$,
it is useful to express the orbit 
using the corotational coordinate system
$(t,X,Y,Z)$ defined as
\begin{align}
 t=t,\quad
 X=x\cos(\Upsilon_{\T{pre}}\lambda)-y\sin(\Upsilon_{\T{pre}}\lambda),\quad
 Y=x\sin(\Upsilon_{\T{pre}}\lambda)+y\cos(\Upsilon_{\T{pre}}\lambda),\quad
Z=z,\label{eq:co-rotational}
\end{align}
where the precession frequency, $\Upsilon_{\T{pre}}$, is defined as
$\Upsilon_{\T{pre}}:=\Upsilon_{\theta}-\Upsilon_{\phi}$.
The right figure in Fig.~\ref{fig:orbit00} is expressed in the corotational coordinate system.
The left figure in Fig.~\ref{fig:orbit01} is a plot of the orbits projected 
onto the $X=0$ plane.
This figure shows that the particle almost remains on
the $Z=Y\tan\theta_{\T{inc}}$ plane. 
In fact, the orbit around a Schwarzschild black hole is exactly on this plane.
On the other hand, 
when the parameter $p$ becomes small and the particle approaches the horizon,
it becomes impossible to define such an approximate orbital plane.
In such a case, it is not appropriate to call the parameter 
$\theta_{\T{inc}}$ the inclination angle. 

The right figure in Fig.~\ref{fig:orbit01} is the
projected image onto the $Z=Y\tan\theta_{\T{inc}}$ plane.
The minimum and maximum distances from
the origin, $r_{\T{min}}$ and $r_{\T{max}}$, are 
$r_{\T{min}} = p/({1+{e}})$ and 
$r_{\T{max}} = p/({1-{e}})$.
In this sense, ${e}$ and ${p}$ are
called the eccentricity and semilatus rectum, respectively.

%%%%%%%%%%%%%%%%%%%%%%%%%%%%%%%%%%%%
\subsection{Homogeneous solutions}\label{sec:homog-solut}
%%%%%%%%%%%%%%%%%%%%%%%%%%%%%%%%%%%

Mano, Suzuki and Takasugi~\cite{Mano:1996vt} (MST) formulated a
method to express
homogeneous solutions of the Teukolsky equation, $R_{\ell m\omega}^{\rm in}$ and
$R_{\ell m\omega}^{\rm up}$, in series of hypergeometric functions 
or Coulomb wave functions. 
Fujita and Tagoshi~\cite{Fujita:2004rb,Fujita:2005rb} applied the MST method 
to the numerical computation of gravitational waves from a particle moving on 
circular, equatorial orbits around a black hole.
To read off the asymptotic amplitudes such as $B^{\T{inc}}$ and 
$B^{\T{ref}}$, we usually have to
numerically integrate the homogeneous Teukolsky equation 
from $r_+$ to a large radius.
In the MST formalism, on the other hand, 
the analytical forms of these asymptotic amplitudes of the homogeneous 
Teukolsky equation are given. We can evaluate the asymptotic amplitude 
very accurately. 
Furthermore, since the convergence of the series of hypergeometric functions 
is fast, we can obtain the homogeneous solutions themselves very accurately. 
In this paper, since we treat eccentric orbits, 
we need to evaluate the homogeneous solutions at many radial points.
Although the convergence of the series of hypergeometric functions or 
Coulomb wave functions is very fast, the computation time required to evaluate
these series is not negligible. 
We thus use the method of successive Taylor series expansions.

We first compute a homogeneous solution at a radius $r$, where
$r_{\rm min}\le r\le r_{\rm max}$, using the series
of hypergeometric functions. This gives a very accurate boundary value
of the solution. 
We then compute the homogeneous solution at $r+h$ using 
the Taylor series expansion around $r$. 
The $n$th derivative of the homogeneous solution at radius $r$
is evaluated from the recurrence relations derived by 
differentiating the homogeneous Teukolsky equation.
The relative error of the Taylor series expansion 
is estimated as $(d^n R^{\T{in}/\T{up}}_{\ell m\omega}(r)/dr^n)h^n/n!/
R^{\T{in}/\T{up}}_{\ell m\omega}(r)$. We adjust $n$ and $h$ to obtain 
a relative error $\sim 10^{-15}$.
Next, we repeat the Taylor series expansion around $r+h$ to obtain the 
homogeneous solution near $r+h$.  
In this way, we obtain $R^{\T{in}/\T{up}}_{\ell m\omega}(r)$ for all radii. 

The Taylor series expansion is used to solve ordinary differential 
equations when high accuracy is required.
The Taylor series method in Ref.~\citen{Corliss-Chang}
is faster than the standard numerical
integration methods when the required accuracy is better than $10^{-6}$. 
We adopt the Taylor series method since we aim to develop a highly accurate code. 
We confirmed that this method gives accurate results and 
is much faster than using the hypergeometric series expansion. 
However, we have not compared the computing time of 
the Taylor series method with that of the standard numerical
integration method by setting the same accuracy. 
The use of the numerical integration method 
together with the MST formalism may be useful for reducing the computation time 
when the required accuracy is not very stringent. 

The spin-weighted spheroidal harmonics, $_{-2}S_{\ell m}^{a\omega}(\theta)$,
are evaluated using a series of Jacobi polynomials. 
The details of the numerical method are described in Ref.~\citen{Fujita:2004rb}.
Although it should be possible to use the Taylor series method
for the spin-weighted spheroidal harmonics, we have not attempted this yet.

%%%%%%%%%%%%%%%%%%%%%%%%%%%%%%%%%%%%%%%%%%%%%%%%%%%%%%%%%%%%%%%%
\section{Results}\label{sec:results}
%%%%%%%%%%%%%%%%%%%%%%%%%%%%%%%%%%%%%%%%%%%%%%%%%%%%%%%%%%%%%%%%

%%%%%%%%%%%%%%%%%%%%%%%%%%
\subsection{Energy spectrum}\label{sec:integration}
%%%%%%%%%%%%%%%%%%%%%%%%%%

In this section, we compute the rates of change of the three constants of motion,
including the Carter constant.

We use the trapezium rule to compute the integral, Eq.~\eqref{eq:zlmkn}. 
This is because the integrand is a periodic function. 
It is well known that if a periodic function is integrated 
over one period, the trapezium rule is a suitable method for 
numerical integration. 

The MST formalism can be applied only
for the case that the frequency is positive, $\omega > 0$.
The mode summation in Eq.~(\ref{eq:dEdt}) is performed using the following equations:
\begin{eqnarray}
\left<\frac{d\C{E}}{dt}\right>_{\rm GW}^\infty
&=&2 \sum_{\ell=2}^{\infty}\sum_{m=-\ell}^{\ell}\sum_{k=-\infty}^{\infty}
\sum_{n=n_0}^{\infty} \left<\frac{d\C{E}_{\ell m k n}}{dt}\right>_{\rm GW}^\infty, 
\\
\left<\frac{d\C{E}}{dt}\right>_{\rm GW}^{\rm H}
&=&2 \sum_{\ell=2}^{\infty}\sum_{m=-\ell}^{\ell}\sum_{k=-\infty}^{\infty}
\sum_{n=n_0}^{\infty} \left<\frac{d\C{E}_{\ell m k n}}{dt}\right>_{\rm GW}^{\rm H}, 
\\
\left<\frac{d\C{E}_{\ell m k n}}{dt}\right>_{\rm GW}^\infty
&\equiv&
\frac{\mu^{-1}}{4\pi\omega^2_{mkn}}
\left|\tilde{Z}^{\infty}_{\ell mkn}\right|^2,
\\
\left<\frac{d\C{E}_{\ell m k n}}{dt}\right>_{\rm GW}^{\rm H}
&\equiv&
\frac{\mu^{-1}\alpha_{\ell mkn}}{4\pi\omega^2_{mkn}}
\left|\tilde{Z}^{\rm H}_{\ell mkn}\right|^2,
\end{eqnarray}
where $n_0$ is the minimum integer such that $m\Upsilon_\phi + k
\Upsilon_\theta + n_0 \Upsilon_r >0$ holds for each $m$ and $k$.
We also define the intermediate modal energy flux for later use:
\begin{eqnarray}
\left<\frac{d\C{E}_{\ell}}{dt}\right>_{\rm GW}^\infty&\equiv&
2 \sum_{m=-\ell}^{\ell}\sum_{k=-\infty}^{\infty}
\sum_{n=n_0}^{\infty} \left<\frac{d\C{E}_{\ell m k n}}{dt}\right>_{\rm GW}^\infty,
\\
\left<\frac{d\C{E}_{\ell m k}}{dt}\right>_{\rm GW}^\infty&\equiv&
2 \sum_{n=n_0}^{\infty} \left<\frac{d\C{E}_{\ell m k n}}{dt}\right>_{\rm GW}^\infty,
\\
\left<\frac{d\C{E}_{\ell}}{dt}\right>_{\rm GW}^{\rm H}&\equiv&
2 \sum_{m=-\ell}^{\ell}\sum_{k=-\infty}^{\infty}
\sum_{n=n_0}^{\infty} \left<\frac{d\C{E}_{\ell m k n}}{dt}\right>_{\rm GW}^{\rm H},
\\
\left<\frac{d\C{E}_{\ell m k}}{dt}\right>_{\rm GW}^{\rm H}&\equiv&
2 \sum_{n=n_0}^{\infty} \left<\frac{d\C{E}_{\ell m k n}}{dt}\right>_{\rm GW}^{\rm H},
\end{eqnarray}

\begin{figure}[htb]
\begin{center}
\includegraphics[scale=0.8]{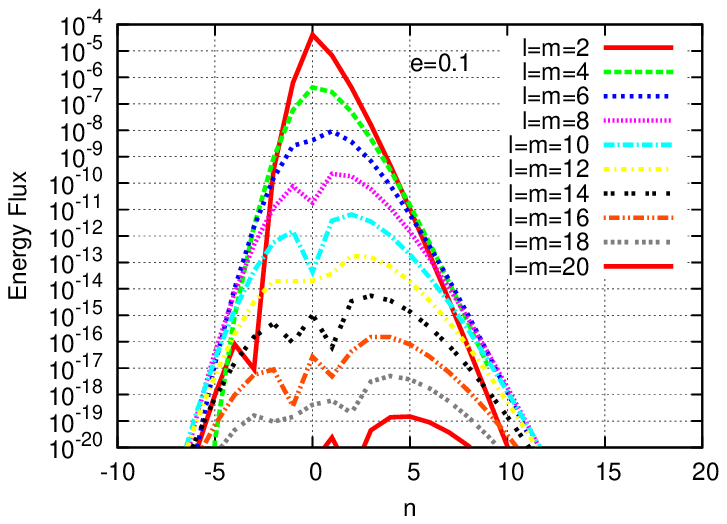}$\quad$
\includegraphics[scale=0.8]{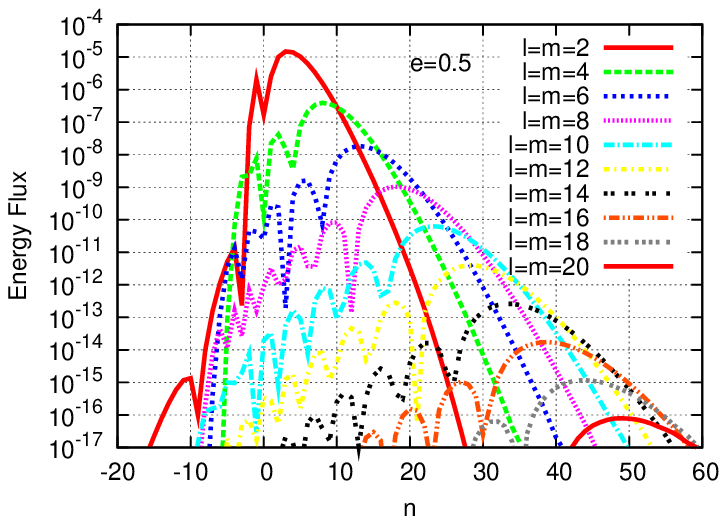}\\
\includegraphics[scale=0.8]{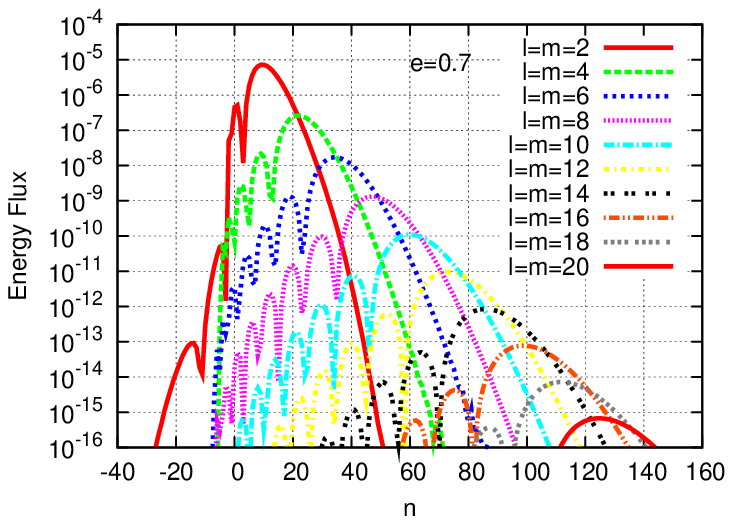}$\quad$
\includegraphics[scale=0.8]{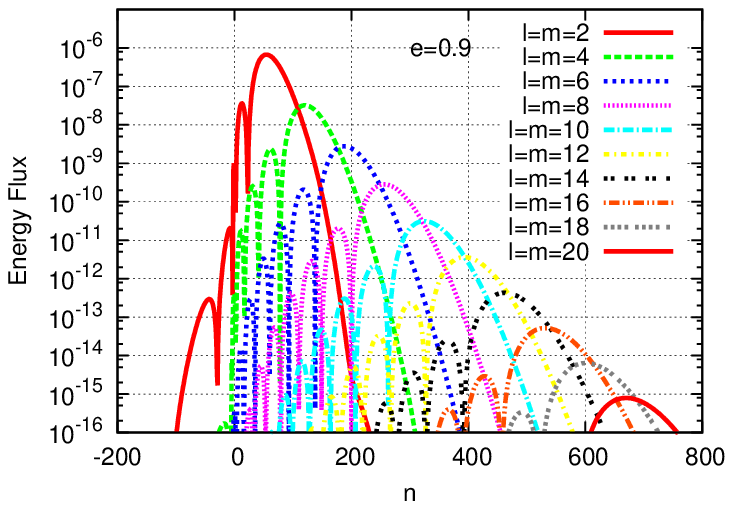}
\caption{Modal energy flux 
$\left<d\C{E}_{\ell m k n}/dt\right>_{\rm GW}^\infty$ at infinity
for $k=0$ and $(p,\theta_{\rm inc})=(10M,20^{\circ})$
in the Schwarzschild case. 
The eccentricities are $0.1$ (top left), 
$0.5$ (top right), $0.7$ (bottom left) and $0.9$ (bottom right).
}\label{fig:nmode1}
\end{center}
\end{figure}

\begin{figure}[htb]
\begin{center}
\includegraphics[scale=0.8]{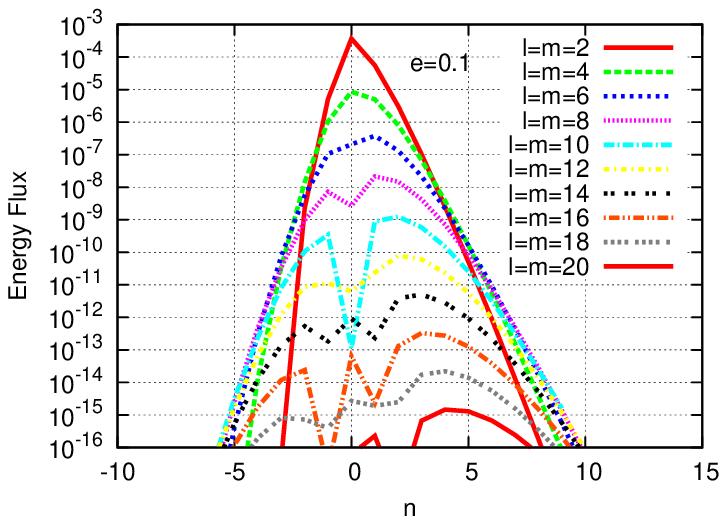}$\quad$
\includegraphics[scale=0.8]{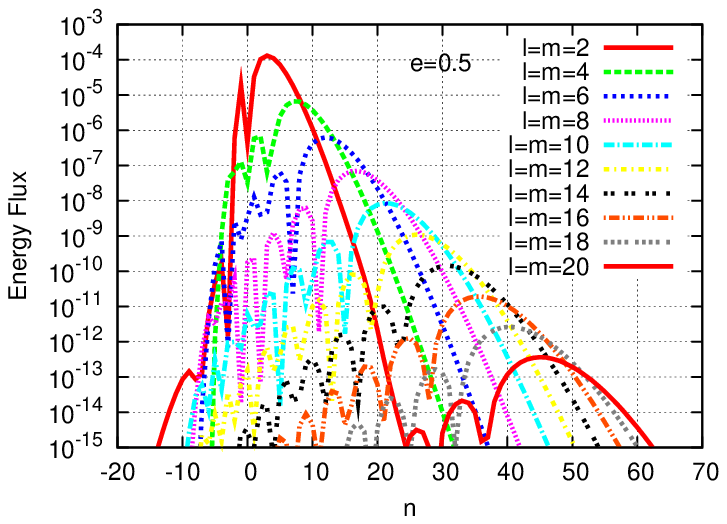}\\
\includegraphics[scale=0.8]{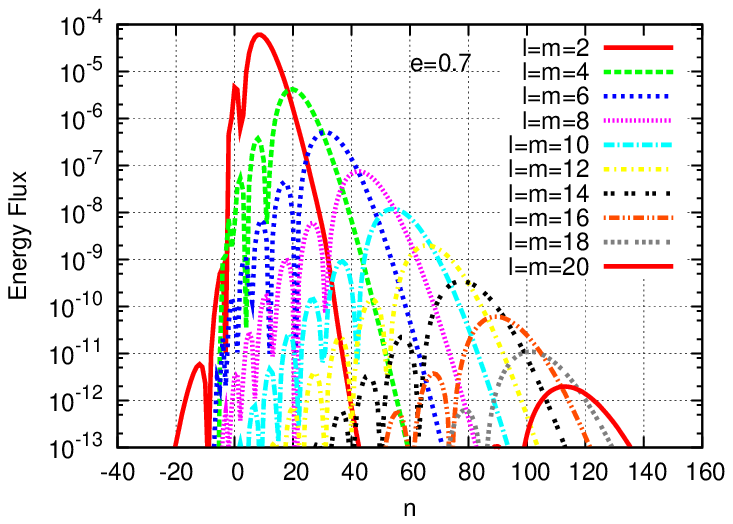}$\quad$
\includegraphics[scale=0.8]{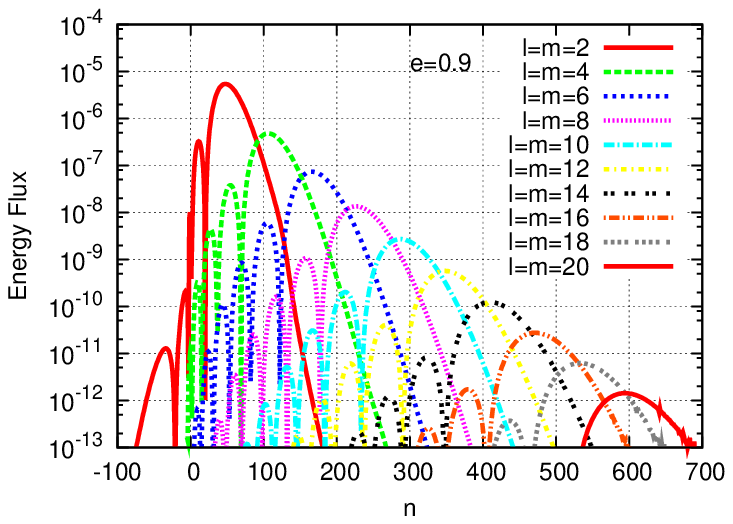}
\caption{Modal energy flux 
$\left<d\C{E}_{\ell m k n}/dt\right>_{\rm GW}^\infty$ at infinity
in the Kerr case ($a=0.9M$) 
for $k=0$ and $(p,\theta_{\rm inc})=(6M,20^{\circ})$.
The eccentricities are $0.1$ (top left), 
$0.5$ (top right), $0.7$ (bottom left) and $0.9$ (bottom right).
}\label{fig:nmode2}
\end{center}
\end{figure}

The behavior of the spectrum in terms of $n$ is complicated.
In Fig.~\ref{fig:nmode1}, we show the modal energy flux at infinity, 
$\left<\left. d\C{E}_{\ell m k n}\right/ dt \right >^{\infty}_{\rm GW}$, 
as functions of $n$ in the Schwarzschild case
and in the cases of $e=0.1$, $0.5$, $0.7$ and $0.9$. 
Figure \ref{fig:nmode2} is the same figure in the Kerr case with $a=0.9M$.
From these figures, we find that the number of peaks of $\left<\left. d\C{E}_{\ell m k n}\right/ dt \right >^{\infty}_{\rm GW}$ is  
roughly $\ell$, and that the value of $n$ at the highest peak of 
$\left<\left. d\C{E}_{\ell m k n}\right/ dt \right >^{\infty}_{\rm GW}$ 
becomes larger as either $\ell$ or the eccentricity $e$ becomes larger.
For example, the location of the peak of  
$\left<\left. d\C{E}_{\ell m k n}\right/ dt \right >^{\infty}_{\rm GW}$ 
is approximately $n=700$ when $e=0.9$, $\ell=m=20$ and $k=0$.
The shapes of the spectra in the Schwarzschild and Kerr cases
are qualitatively very similar. 

In practical computations, we have to truncate the mode summation. 
We set the target accuracy of the mode summation with respect to $n$ and $k$ to be $10^{-10}$.
It is useful if we know the location of the highest 
peak of $\left<\left. d\C{E}_{\ell m k n}\right/ dt \right >^{\infty}_{\rm GW}$
before the computation to reduce the computational time. 
However, it is difficult to derive an analytical expression for such a location. 
In this work, we adopt the following procedure to determine the range of the $n$-mode
summation. 
First, we compute $\left<\left. d\C{E}_{\ell m k n}\right/ dt \right >^{\infty}_{\rm GW}$ for $\ell=m=2$ and $k=0$ 
for a wide range of $n$ starting from $n=n_0$ to obtain 
sufficient coverage of the $n$-mode. 
In this computation, we obtain the location of the peak $n=n_p$ for $\ell=m=2$ and $k=0$. 
For other $\ell=2$ modes,  since it is expected that 
the peak value of $n$ is not significantly different from that for $\ell=m=2$ and $k=0$,
we first search for the peak near $n_p$. 
We then compute $\left<\left. d\C{E}_{\ell m k n}\right/ dt \right >^{\infty}_{\rm GW}$ starting from the new peak of $n$, 
and sum $\left<\left. d\C{E}_{\ell m k n}\right/ dt \right >^{\infty}_{\rm GW}$ until the total flux
converges with an accuracy of $10^{-10}$. 
For modes $(\ell,m,k)$ with $\ell>2$, we basically repeat the above procedure. 
We search for the peak near the 
peak for $(\ell,m,k)=(\ell-1,\ell-1,0)$. 
Starting from this location of $n$, we compute and sum $\left<\left. d\C{E}_{\ell m k n}\right/ dt \right >^{\infty}_{\rm GW}$ until
the total flux converges.

\begin{figure}[htb]
\begin{center}
\includegraphics[scale=1.0]{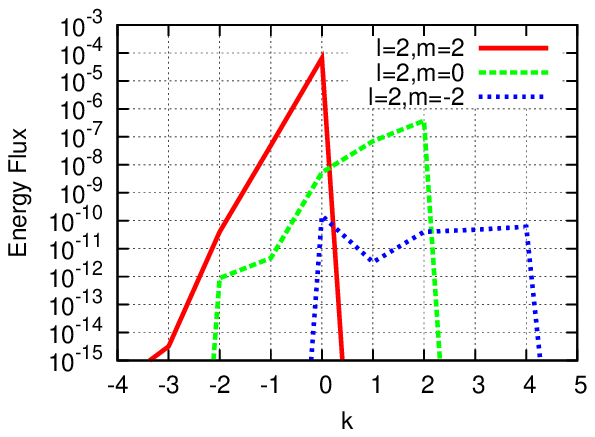}
$\quad$
\includegraphics[scale=1.0]{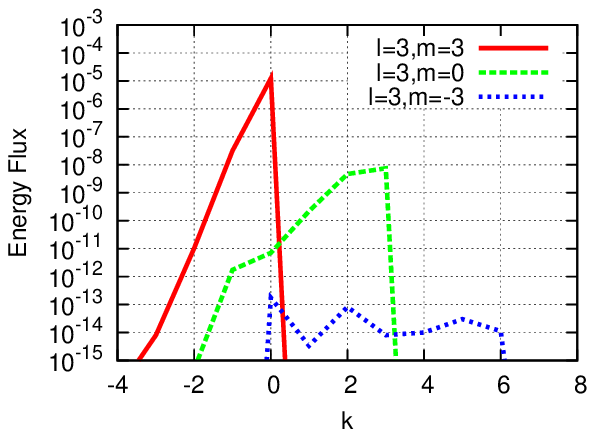}\\
\includegraphics[scale=1.0]{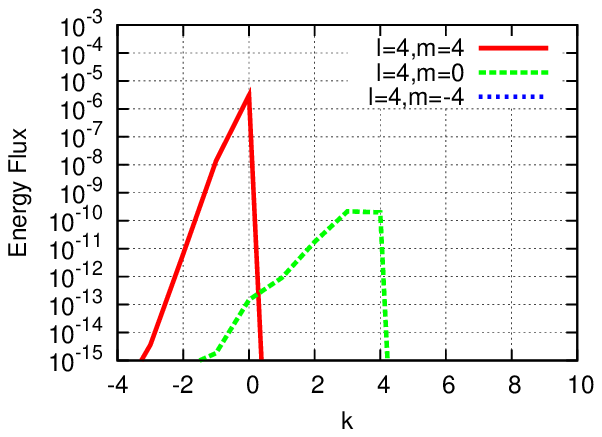}
$\quad$
\includegraphics[scale=1.0]{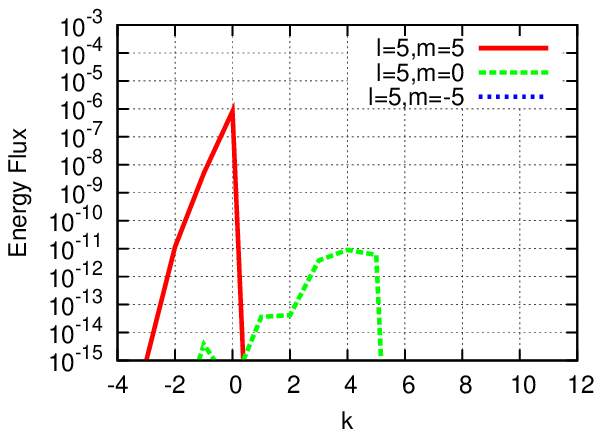}
\end{center}
\caption{
Modal energy flux 
$\left<d\C{E}_{\ell m k}/dt\right>_{\rm GW}^\infty$ at infinity
in the Schwarzschild case. 
$(p,e,\theta_{\rm inc})=(10M,0.7,20^{\circ})$. 
$\ell=2$ (top left), $\ell=3$ (top right), $\ell=4$ (bottom left)
and $\ell=5$ (bottom right). 
These figures show that the peaks are located at approximately
$k=\ell-m$.
}\label{fig:kmode1}
\end{figure}

\begin{figure}[htb]
\begin{center}
\includegraphics[scale=1.0]{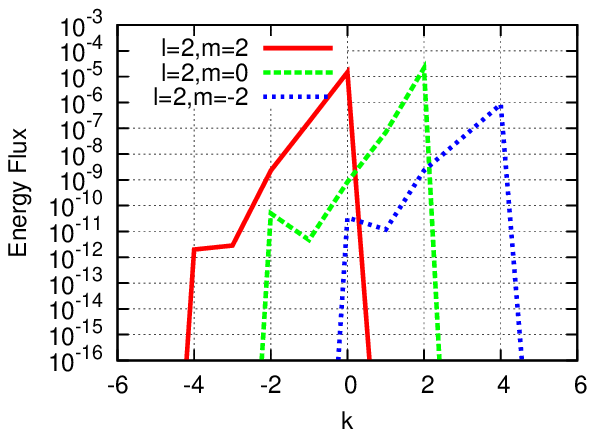}
$\quad$
\includegraphics[scale=1.0]{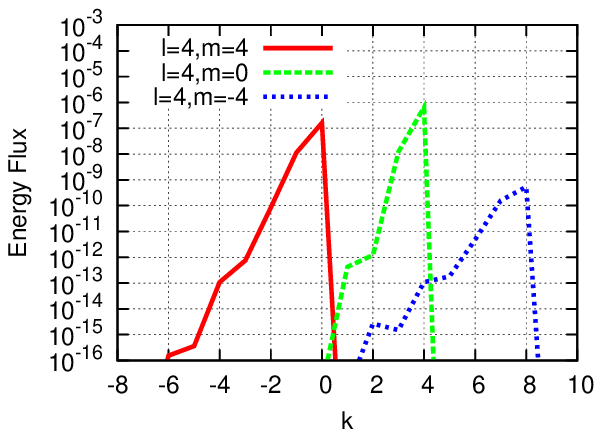}\\
\includegraphics[scale=1.0]{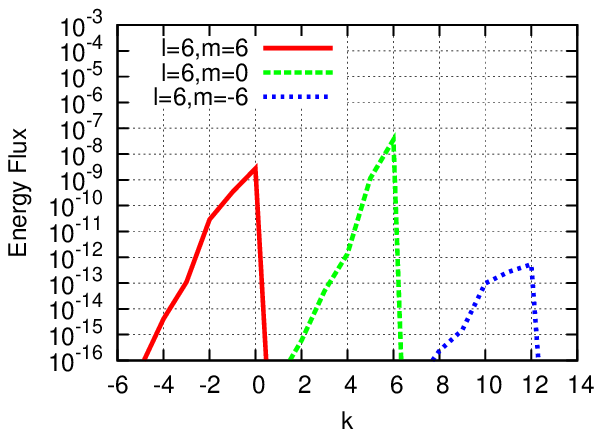}
$\quad$
\includegraphics[scale=1.0]{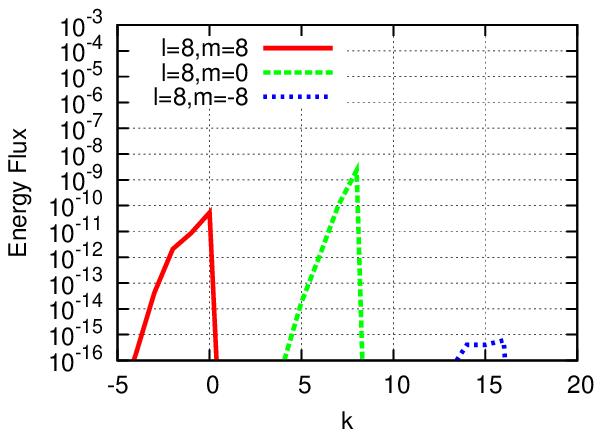}
\caption{Modal energy flux 
$\left<d\C{E}_{\ell m k}/dt\right>_{\rm GW}^\infty$ at infinity
in the Schwarzschild case. 
$(p,e,\theta_{\rm inc})=(10M,0.7,70^{\circ})$. 
$\ell=2$ (top left), $\ell=4$ (top right), $\ell=6$ (bottom left)
and $\ell=8$ (bottom right). 
These figures show that the peaks are located at approximately $k=\ell-m$.
}\label{fig:kmode2}
\end{center}
\end{figure}

\begin{figure}[htb]
\begin{center}
\includegraphics[scale=1.0]{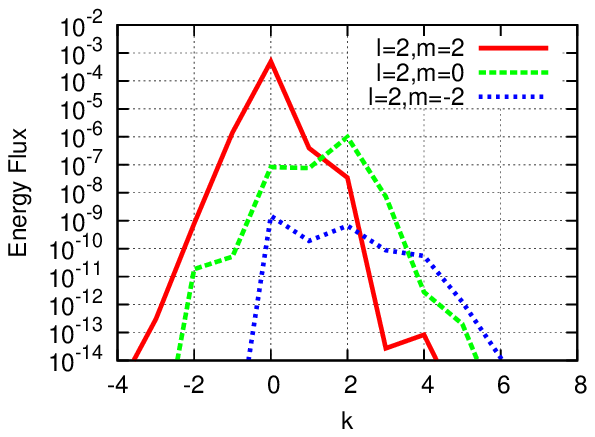}
$\quad$
\includegraphics[scale=1.0]{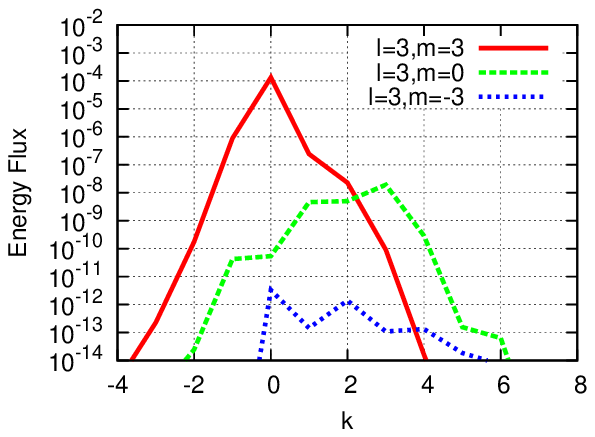}\\
\includegraphics[scale=1.0]{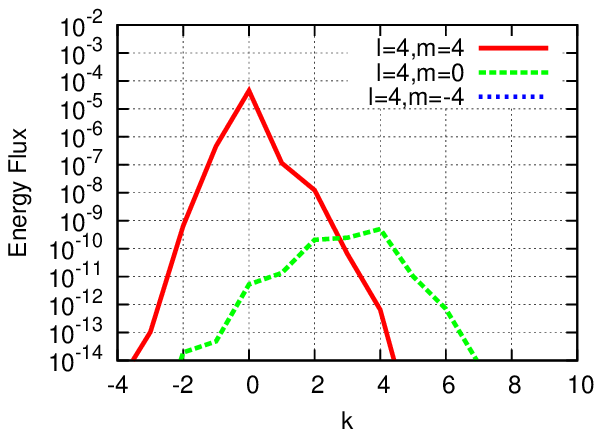}
$\quad$
\includegraphics[scale=1.0]{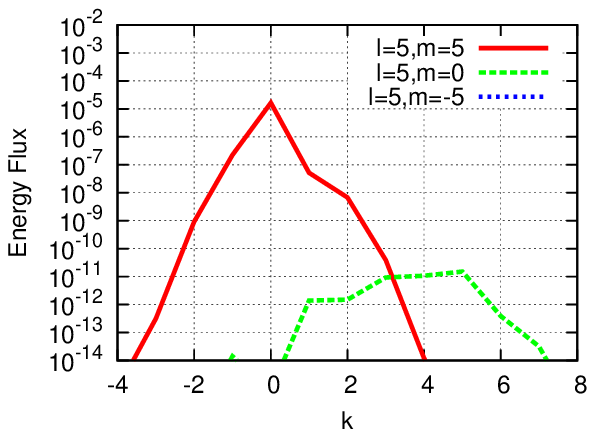}
\end{center}
\caption{
Modal energy flux 
$\left<d\C{E}_{\ell m k}/dt\right>_{\rm GW}^\infty$ at infinity
in the Kerr case ($a=0.9M$). 
$(p,e,\theta_{\rm inc})=(6M,0.7,20^{\circ})$. 
$\ell=2$ (top left), $\ell=3$ (top right), $\ell=4$ (bottom left)
and $\ell=5$ (bottom right). 
These figures show that the peaks are located at approximately $k=\ell-m$.
}\label{fig:kmode3}
\end{figure}

\begin{figure}[htb]
\begin{center}
\includegraphics[scale=1.0]{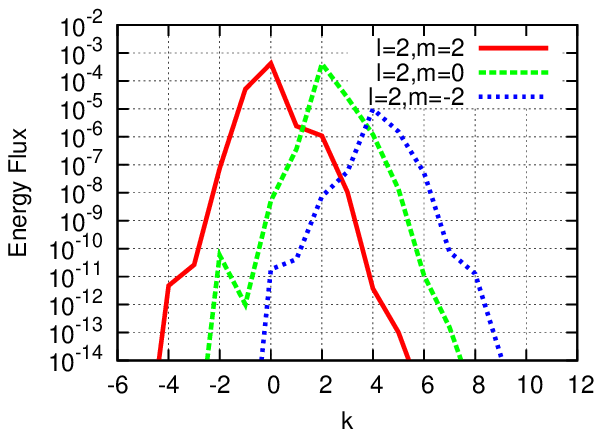}
$\quad$
\includegraphics[scale=1.0]{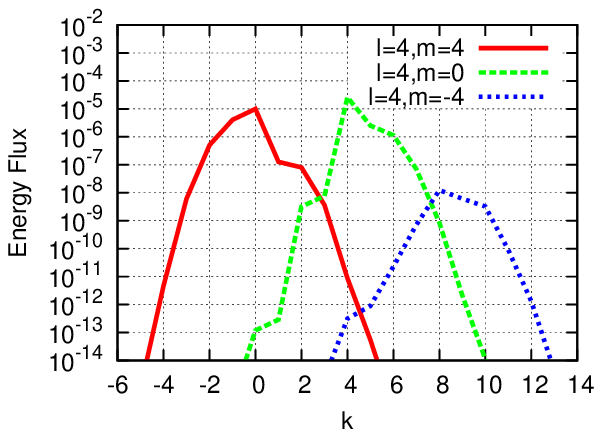}\\
\includegraphics[scale=1.0]{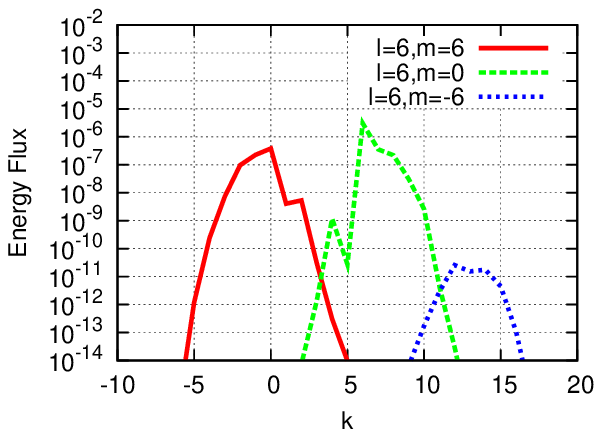}
$\quad$
\includegraphics[scale=1.0]{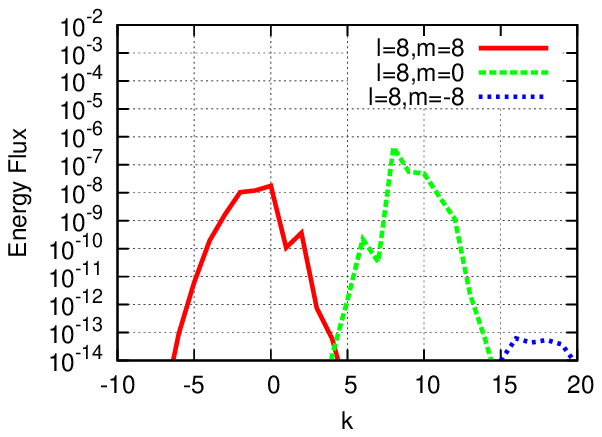}
\caption{Modal energy flux 
$\left<d\C{E}_{\ell m k}/dt\right>_{\rm GW}^\infty$ at infinity
in the Kerr case ($a=0.9M$). 
$(p,e,\theta_{\rm inc})=(6M,0.7,80^{\circ})$. 
$\ell=2$ (top left), $\ell=4$ (top right), $\ell=6$ (bottom left)
and $\ell=8$ (bottom right). 
These figures show that the peaks are located at approximately $k=\ell-m$.
}\label{fig:kmode4}
\end{center}
\end{figure}

We show the spectrum of 
$\left<\left. d\C{E}_{\ell m k}\right/ dt \right >^{\infty}_{\rm GW}$ 
as functions of $k$
in Figs.~\ref{fig:kmode1} and \ref{fig:kmode2} in the Schwarzschild case, 
and in Figs.~\ref{fig:kmode3} and \ref{fig:kmode4} in the Kerr case with $a=0.9M$. 
Figures \ref{fig:kmode1} and \ref{fig:kmode3} show the spectrum for the inclination angle 
$\theta_{\rm inc}=20^\circ$,
and Figs.~\ref{fig:kmode2} and \ref{fig:kmode4} show the spectrum for the high inclination angles 
$\theta_{\rm inc}=70^\circ$ and $80^\circ$, respectively. 
We find that the peak of the $k$-mode usually exists at approximately $\ell-m$ except for
the $m<0$ modes in the case of a low inclination angle. 
We also find that when the inclination angle is $20^\circ$,
the peak value of 
the modal energy flux becomes smaller when $m$ changes from $\ell$ to $-\ell$.
On the other hand, when the inclination angle is large, 
i.e., $\theta_{\rm inc}=70^\circ$ or $80^\circ$,
the peak value is largest when $m=0$.
Similarly to the $n$-mode, the difference between the Schwarzschild and Kerr cases 
is not very large, but the spectra of the Kerr cases are broader than those of the Schwarzschild cases.
By taking this behavior into account, 
it is straightforward to 
truncate the $k$-mode summation, which guarantees the accuracy of the total flux of $10^{-10}$.

\begin{figure}[htb]
\begin{center}
\includegraphics[scale=0.8]{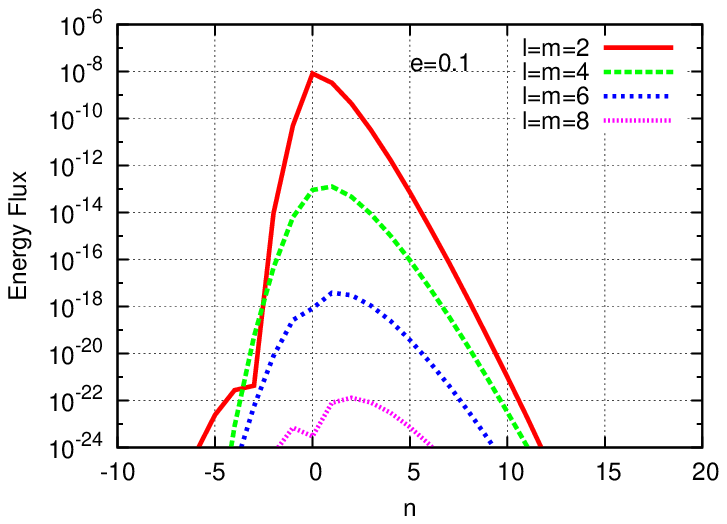}$\quad$
\includegraphics[scale=0.8]{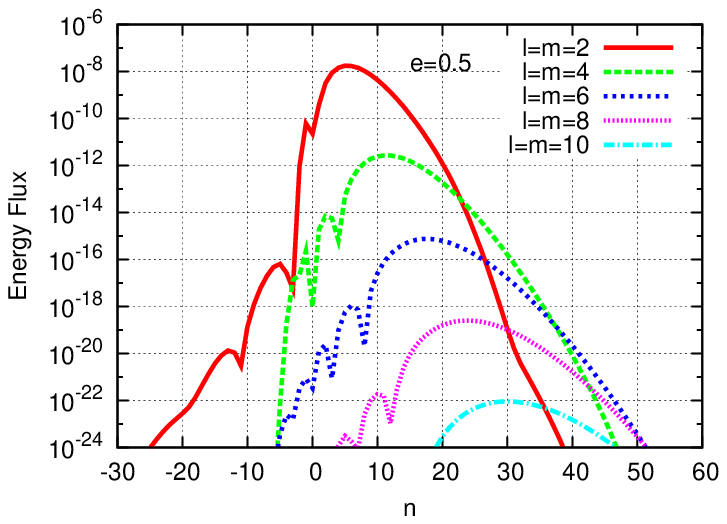}\\
\includegraphics[scale=0.8]{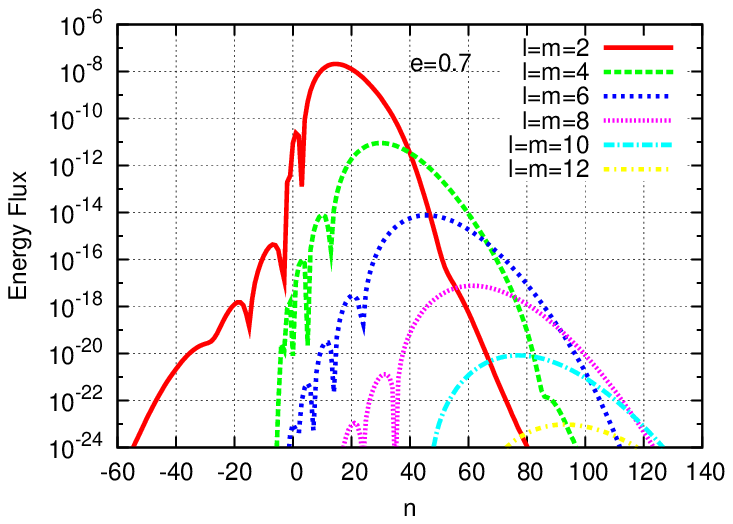}$\quad$
\includegraphics[scale=0.8]{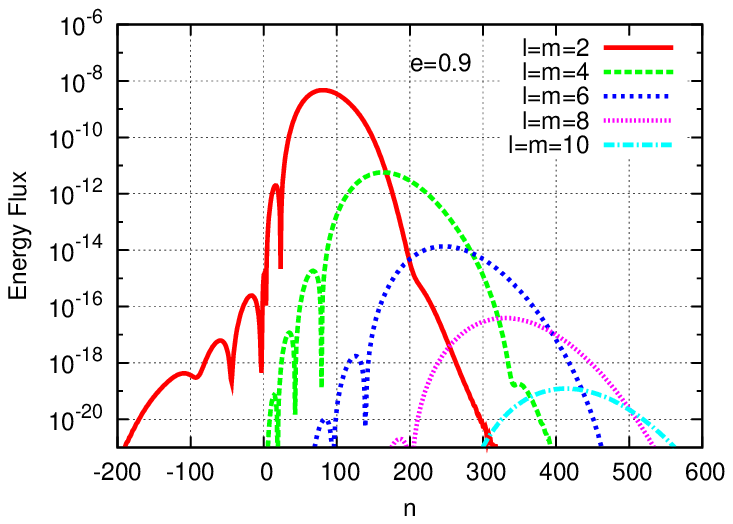}
\caption{Modal energy flux 
$\mid\left<d\C{E}_{\ell m k n}/dt\right>_{\rm GW}^{\rm H}\mid$ at the horizon
for $k=0$ and  $(p,\theta_{\rm inc})=(10M,20^{\circ})$
in the Schwarzschild case. 
The eccentricities are $0.1$ (top left), 
$0.5$ (top right), $0.7$ (bottom left) and $0.9$ (bottom right).
}\label{fig:nmodeH1}
\end{center}
\end{figure}

\begin{figure}[htb]
\begin{center}
\includegraphics[scale=0.8]{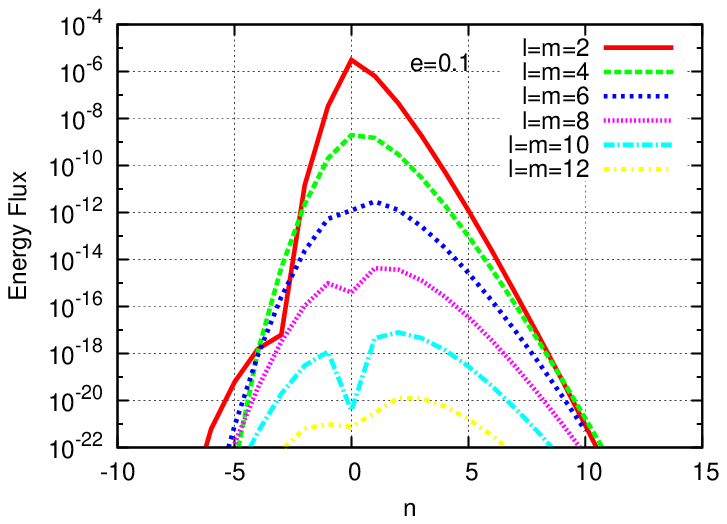}$\quad$
\includegraphics[scale=0.8]{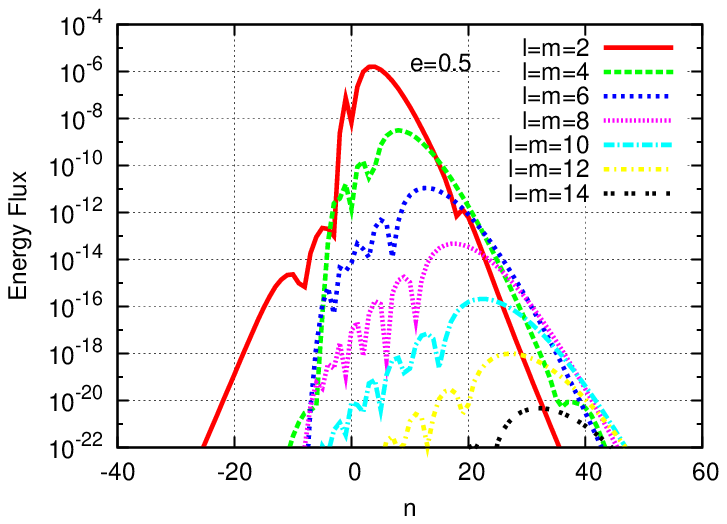}\\
\includegraphics[scale=0.8]{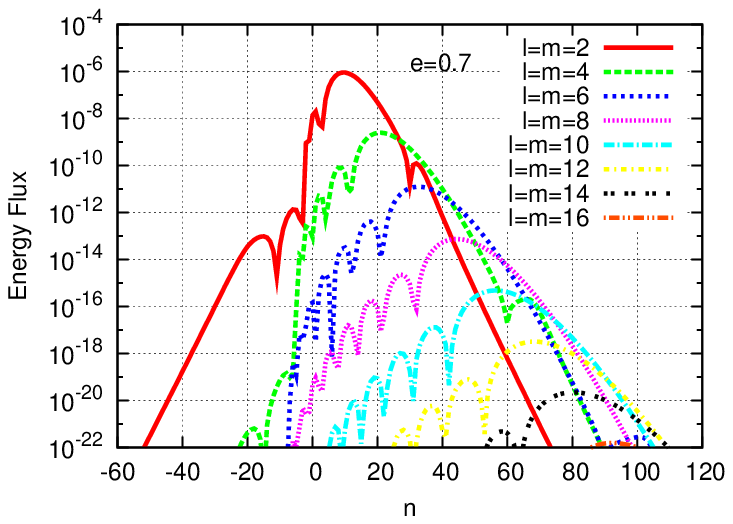}$\quad$
\includegraphics[scale=0.8]{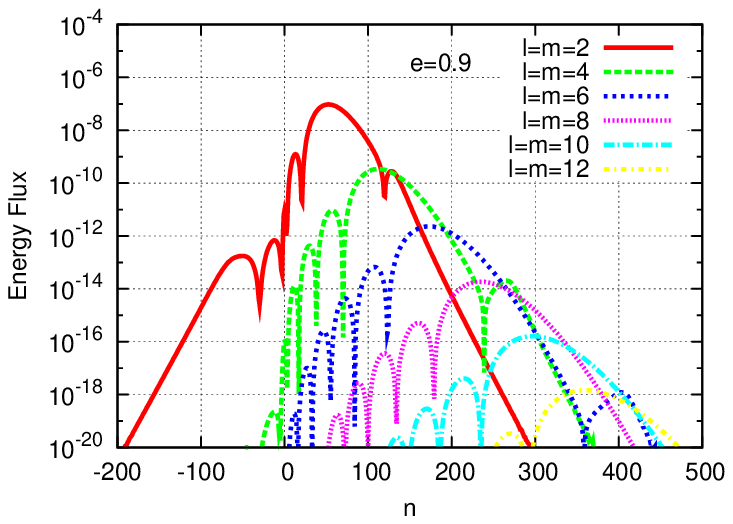}
\caption{Modal energy flux 
$\mid\left<d\C{E}_{\ell m k n}/dt\right>_{\rm GW}^{\rm H}\mid$ at the horizon
in the Kerr case ($a=0.9M$) 
for $k=0$ and  $(p,\theta_{\rm inc})=(6M,20^{\circ})$.
The eccentricities are $0.1$ (top left), 
$0.5$ (top right), $0.7$ (bottom left) and $0.9$ (bottom right).
}\label{fig:nmodeH2}
\end{center}
\end{figure}

\begin{figure}[htb]
\begin{center}
\includegraphics[scale=1.0]{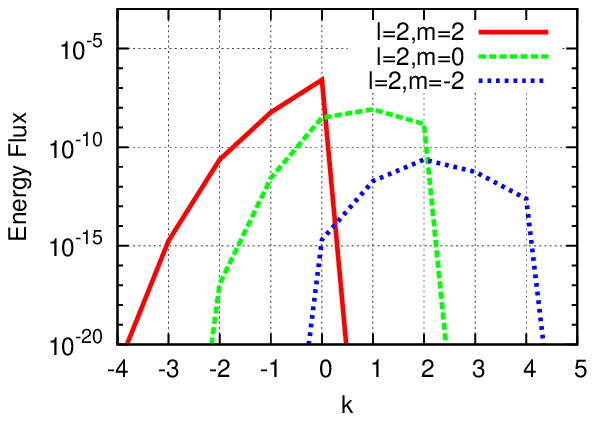}
$\quad$
\includegraphics[scale=1.0]{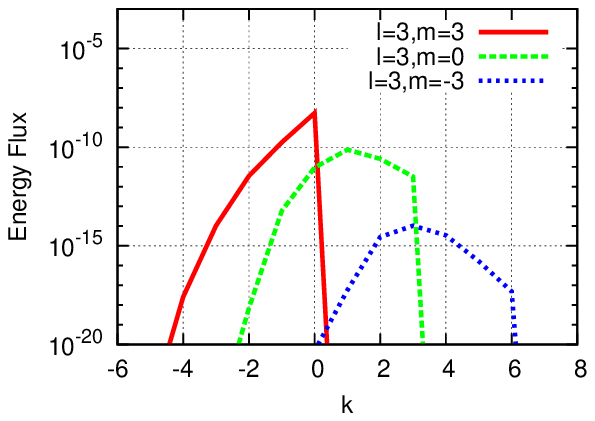}\\
\includegraphics[scale=1.0]{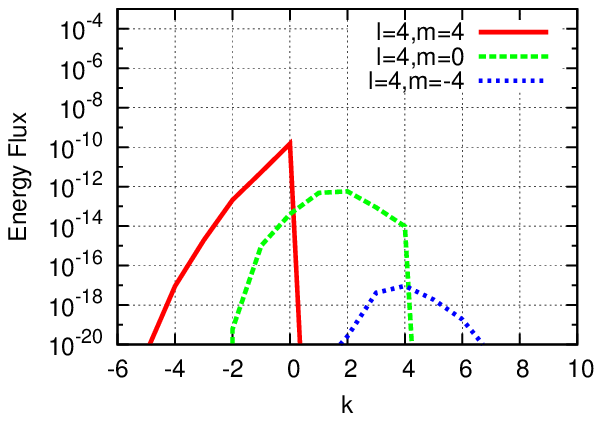}
$\quad$
\includegraphics[scale=1.0]{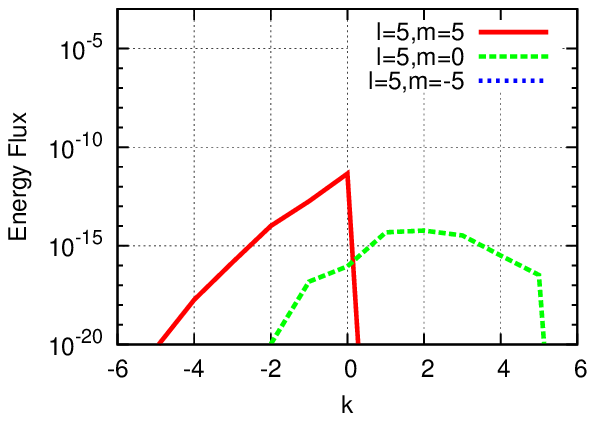}
\end{center}
\caption{
Modal energy flux 
$\mid\left<d\C{E}_{\ell m k}/dt\right>_{\rm GW}^{\rm H}\mid$ at the horizon
in the Schwarzschild case. 
$(p,e,\theta_{\rm inc})=(10M,0.7,20^{\circ})$. 
$\ell=2$ (top left), $\ell=3$ (top right), $\ell=4$ (bottom left)
and $\ell=5$ (bottom right). 
These figures show that the peaks are located at approximately $k=\ell-m$.
}\label{fig:kmodeH1}
\end{figure}

\begin{figure}[htb]
\begin{center}
\includegraphics[scale=1.0]{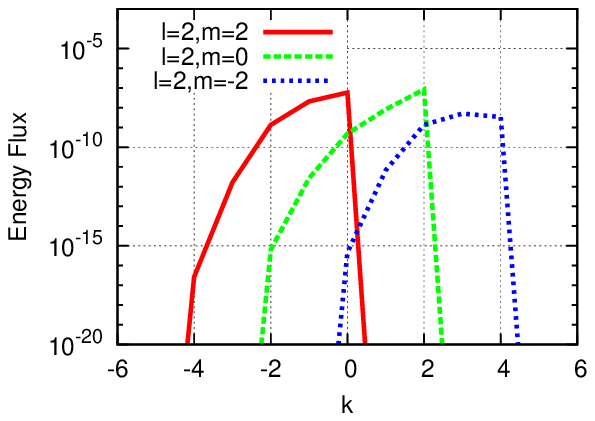}
$\quad$
\includegraphics[scale=1.0]{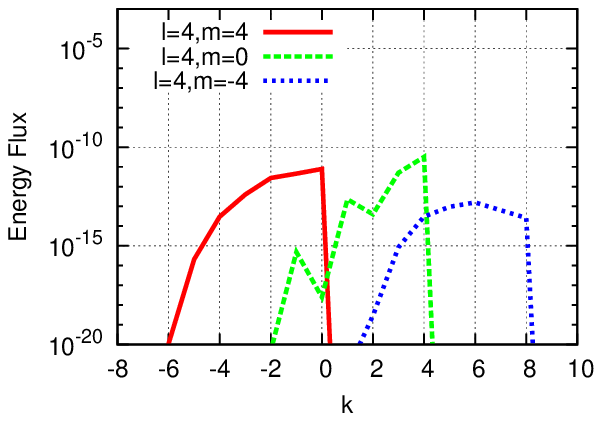}\\
\includegraphics[scale=1.0]{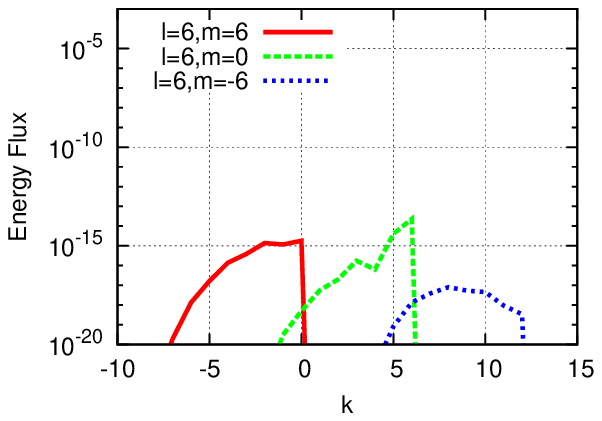}
$\quad$
\includegraphics[scale=1.0]{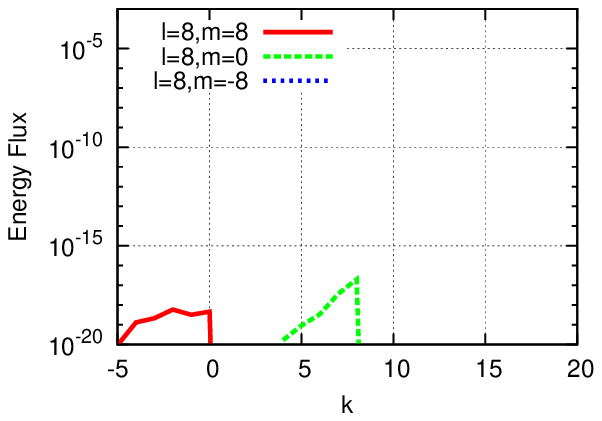}
\caption{Modal energy flux 
$\mid\left<d\C{E}_{\ell m k}/dt\right>_{\rm GW}^{\rm H}\mid$ at the horizon
in the Schwarzschild case. 
$(p,e,\theta_{\rm inc})=(10M,0.7,70^{\circ})$. 
$\ell=2$ (top left), $\ell=4$ (top right), $\ell=6$ (bottom left)
and $\ell=8$ (bottom right). 
These figures show that the peaks are located at approximately $k=\ell-m$.
}\label{fig:kmodeH2}
\end{center}
\end{figure}

\begin{figure}[htb]
\begin{center}
\includegraphics[scale=1.0]{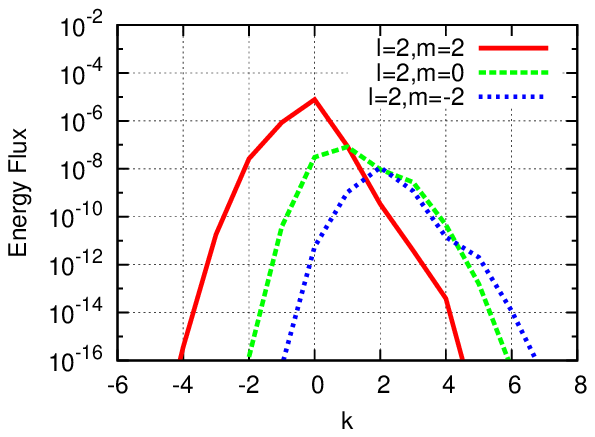}
$\quad$
\includegraphics[scale=1.0]{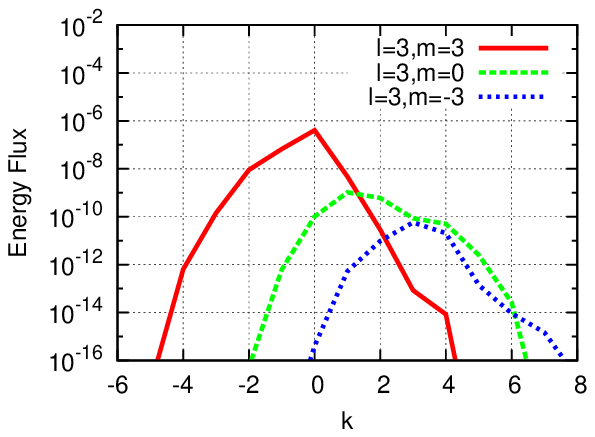}\\
\includegraphics[scale=1.0]{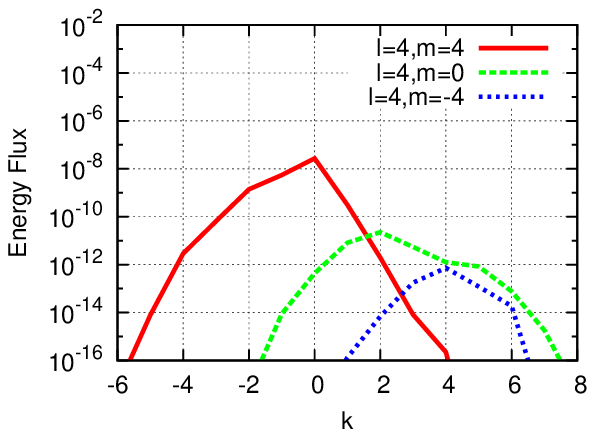}
$\quad$
\includegraphics[scale=1.0]{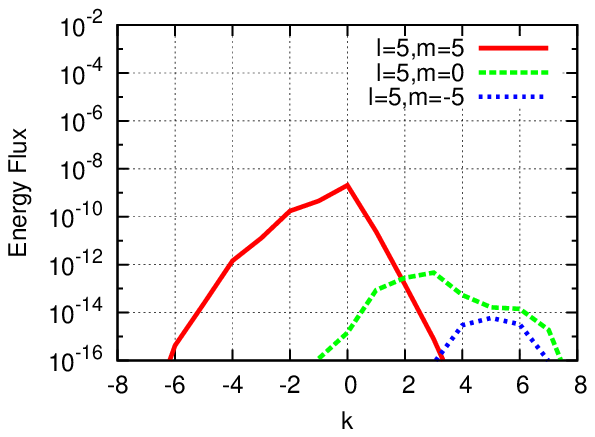}
\end{center}
\caption{
Modal energy flux 
$\mid\left<d\C{E}_{\ell m k}/dt\right>_{\rm GW}^{\rm H}\mid$ at the horizon
in the Kerr case ($a=0.9M$). 
$(p,e,\theta_{\rm inc})=(6M,0.7,20^{\circ})$. 
$\ell=2$ (top left), $\ell=3$ (top right), $\ell=4$ (bottom left)
and $\ell=5$ (bottom right). 
These figures show that the peaks are located at approximately $k=\ell-m$.
}\label{fig:kmodeH3}
\end{figure}

\begin{figure}[htb]
\begin{center}
\includegraphics[scale=1.0]{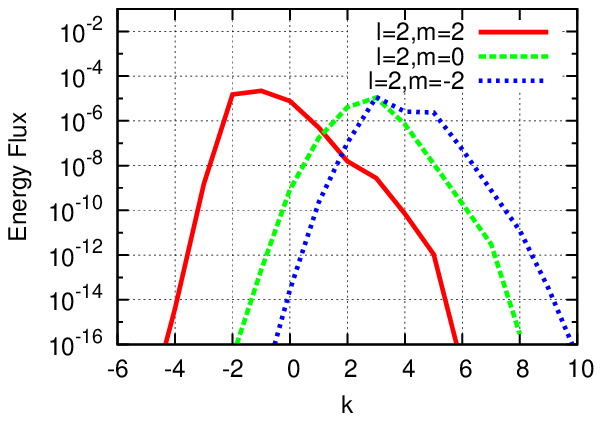}
$\quad$
\includegraphics[scale=1.0]{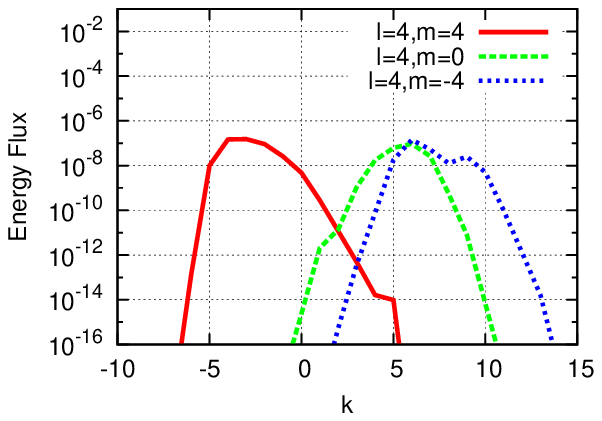}\\
\includegraphics[scale=1.0]{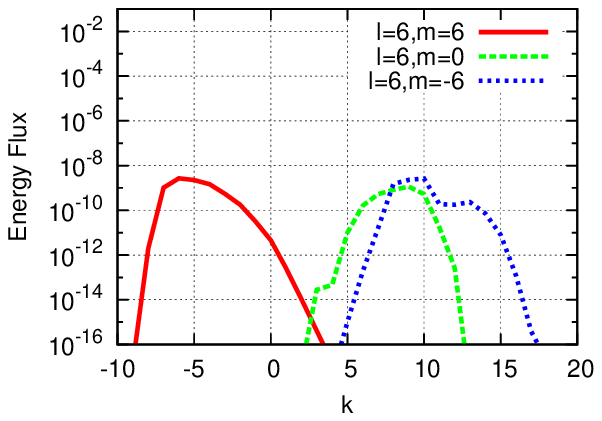}
$\quad$
\includegraphics[scale=1.0]{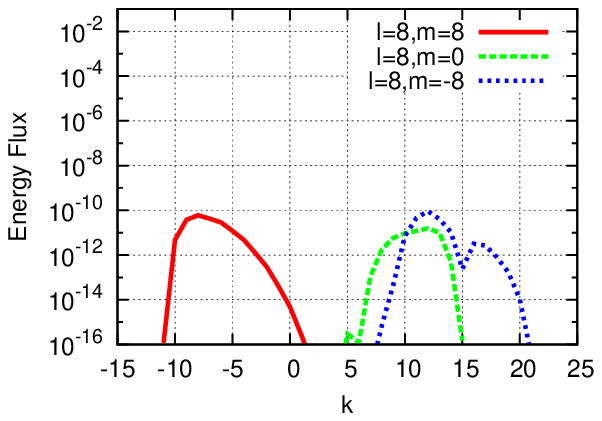}
\caption{Modal energy flux 
$\mid\left<d\C{E}_{\ell m k}/dt\right>_{\rm GW}^{\rm H}\mid$ at the horizon
in the Kerr case ($a=0.9M$). 
$(p,e,\theta_{\rm inc})=(6M,0.7,80^{\circ})$. 
$\ell=2$ (top left), $\ell=4$ (top right), $\ell=6$ (bottom left)
and $\ell=8$ (bottom right). 
These figures show that the peaks are located at approximately $k=\ell-m$.
}\label{fig:kmodeH4}
\end{center}
\end{figure}

In Figs.~\ref{fig:nmodeH1} and \ref{fig:nmodeH2}, we show the modal energy flux at the horizon, 
$\left<\left. d\C{E}_{\ell m k n}\right/ dt \right >_{\rm GW}^{\rm H}$, 
as functions of $n$. 
In Figs.~\ref{fig:kmodeH1}--\ref{fig:kmodeH4}, 
we show the spectrum of 
$\left<\left. d\C{E}_{\ell m k}\right/ dt \right >_{\rm GW}^{\rm H}$ 
as functions of $k$. 
We find that the shape of the energy spectrum at the horizon is very similar to that 
at infinity. We thus follow the above procedure to determine the range of summation of $k$ and $n$
for the energy spectrum at the horizon. 

For the $m$-mode, 
we compute all $m$-modes from $\ell$ to $-\ell$. 
For $\ell$-mode, we set the maximum of $\ell$ to be $\ell_{\rm max}=20$. 
This value is chosen to reduce computation time.
Since the energy flux for each $\ell$-mode
monotonically decreases with increasing $\ell$,
the relative error of the total flux due to this truncation 
is estimated as 
\begin{eqnarray}
&&\Delta_{(\ell_{\rm max})}^{\infty,{\rm H}}\equiv \left<\frac{d\C{E}_{\ell_{\rm max}}}{dt}\right>_{\rm GW}
^{\infty,{\rm H}}/\displaystyle\sum_{\ell=2}^{\ell_{\rm max}}\left<\frac{d\C{E}_{\ell}}{dt}\right>_{\rm GW}^{\infty,{\rm H}}.
\end{eqnarray}

%%%%%%%%%%%%%%%%%%%%%%%%%%%%%%%%%%%%%%%%%%%%%%%%%%%%%%%%%%%%%%%%%%%%%%%%%%%
\subsection{Rates of change of the three constants of motion}\label{sec:high-eccentr-orbits}
%%%%%%%%%%%%%%%%%%%%%%%%%%%%%%%%%%%%%%%%%%%%%%%%%%%%%%%%%%%%%%%%%%%%%%%%%%%

We first show the results of the rate of change of energy and
the energy spectrum in the case of a Schwarzschild black hole. 
Indeed, the Schwarzschild cases are useful for verifying the accuracy 
of our code. In these cases, the total energy flux 
after summing the $\ell$-, $m$-, $k$- and $n$-modes
is independent of the inclination angle. 
We verify this property by computing the rate of change of energy 
when the particle moves on the plane with an inclination 
angle from the equatorial plane around a Schwarzschild black hole.
In Table \ref{tab:Schwarz_flux}, we show the data for different inclination angles 
and eccentricities. We find that, 
even if the inclination angle is changed, 
$\left<\left. d\C{E}\right/ dt \right >^{\infty}$
remains within the accuracy of $10^{-8}-10^{-10}$. 
In this table, truncation errors $\Delta_{(\ell_{\rm max})}^\infty$ are shown
in square brackets. $\Delta_{(\ell_{\rm max})}^\infty$ is similar to 
the truncation errors of $n$- and $k$-modes, $10^{-10}$. 
Thus, the errors in Table \ref{tab:Schwarz_flux} are consistent with
the truncation errors of the summation of the $\ell$-, $n$- and $k$-mode.

In Table \ref{tab:Schwarz_fluxH}, we show similar results for the horizon flux. 
In these cases, the truncation error from the $\ell$-mode summation is negligible. 
The relative error in Table \ref{tab:Schwarz_fluxH} is due to the truncation of the
$n$- and $k$-mode summation.

\begin{table}[htbp]
\begin{center}
\caption{Time-averaged rates of change of the energy of a particle 
due to the emission of gravitational waves to infinity for the case of
a Schwarzschild black hole.
In this table, the orbital radius is $10M$.
We compare the result for the equatorial plane with that for
a nonequatorial plane.
Truncation errors $\Delta_{(\ell_{\rm max})}^\infty$ are shown
in square brackets. 
Here we set $\ell_{\T{max}}=20$.
These results show that the error  of $\left<\left. d\C{E}\right/ dt \right >^{\infty}$
is consistent with the truncation error of the mode summation. }
\begin{tabular}{c c c c||c|c}
\hline \hline
  $\ a/M\ $  & $\ \ p/M\ \ $ & $\ \ e\ \ $ &$ \theta_{\T{inc}}$
  & $\ \left<\left. d\C{E}\right/ dt \right >^{\infty}\ $
  & Relative error \\\hline
0&10&0.1& 0$^{\circ}$& $-6.31752474720\times 10^{-5}$ & $[10^{-11}]$\\
0&10&0.1&20$^{\circ}$& $-6.31752474730\times 10^{-5}$ & 1.6$\times 10^{-11}$ \\
0&10&0.1&45$^{\circ}$& $-6.31752474742\times 10^{-5}$ & 3.5$\times 10^{-11}$ \\
0&10&0.1&70$^{\circ}$& $-6.31752474665\times 10^{-5}$ & 8.7$\times 10^{-11}$ \\
  \hline
0 &10&0.5& 0$^{\circ}$&$-9.27335011503\times 10^{-5}$& $[10^{-11}]$\\
0 &10&0.5&20$^{\circ}$&$-9.27335011442\times 10^{-5}$& 6.6$\times 10^{-11}$ \\
0 &10&0.5&45$^{\circ}$&$-9.27335011373\times 10^{-5}$& 1.4$\times 10^{-10}$\\
0 &10&0.5&70$^{\circ}$&$-9.27335011191\times 10^{-5}$& 3.4$\times 10^{-10}$\\
  \hline
0&10&0.7& 0$^{\circ}$&$-9.46979134409\times 10^{-5}$& $[10^{-9}]$\\
0&10&0.7&20$^{\circ}$&$-9.46979134028\times 10^{-5}$&4.0$\times 10^{-10}$\\
0&10&0.7&45$^{\circ}$&$-9.46979133931\times 10^{-5}$&5.0$\times 10^{-10}$\\
0&10&0.7&70$^{\circ}$&$-9.46979131018\times 10^{-5}$&3.6$\times 10^{-9}$\\
  \hline
0&10&0.9& 0$^{\circ}$&$-4.19426469206\times 10^{-5}$& $[10^{-8}]$\\
0&10&0.9&20$^{\circ}$&$-4.19426468442\times 10^{-5}$&1.8$\times 10^{-9}$\\
0&10&0.9&45$^{\circ}$&$-4.19426468407\times 10^{-5}$&1.9$\times 10^{-9}$\\
0&10&0.9&70$^{\circ}$&$-4.19426437158\times 10^{-5}$&7.6$\times 10^{-8}$\\
\hline \hline
\end{tabular}
\label{tab:Schwarz_flux}
\end{center}
\end{table}

\begin{table}[htbp]
\begin{center}
\caption{Time-averaged rates of change of the energy of a particle 
due to the emission of gravitational waves to the horizon for the case of 
a Schwarzschild black hole.
In this table, the orbital radius is $10M$.
We compare the result for the equatorial plane with that for
a nonequatorial plane.
Truncation errors $\Delta_{(\ell_{\rm max})}^{\rm H}$ are shown
in square brackets. 
Here we set $\ell_{\T{max}}=20$.
}
\begin{tabular}{c c c c||c|c}
\hline \hline
  $\ a/M\ $  & $\ \ p/M\ \ $ & $\ \ e\ \ $ &$ \theta_{\T{inc}}$
  & $\ \left<\left. d\C{E}\right/ dt \right >^{\rm H}\ $
  & Relative error \\\hline
0&10&0.1& 0$^{\circ}$& $-1.53365819445\times 10^{-8}$ & $[10^{-35}]$\\
0&10&0.1&20$^{\circ}$& $-1.53365819444\times 10^{-8}$ & 6.5$\times 10^{-12}$ \\
0&10&0.1&45$^{\circ}$& $-1.53365819443\times 10^{-8}$ & 1.3$\times 10^{-11}$ \\
0&10&0.1&70$^{\circ}$& $-1.53365819444\times 10^{-8}$ & 6.5$\times 10^{-12}$ \\
  \hline
0&10&0.5& 0$^{\circ}$&$-1.41298859260\times 10^{-7}$& $[10^{-31}]$\\
0&10&0.5&20$^{\circ}$&$-1.41298859246\times 10^{-7}$& 9.9$\times 10^{-11}$ \\
0&10&0.5&45$^{\circ}$&$-1.41298858862\times 10^{-7}$& 2.8$\times 10^{-9}$\\
0&10&0.5&70$^{\circ}$&$-1.41298859240\times 10^{-7}$& 1.4$\times 10^{-10}$\\
  \hline
0&10&0.7& 0$^{\circ}$&$-3.55415030114\times 10^{-7}$& $[10^{-26}]$\\
0&10&0.7&20$^{\circ}$&$-3.55415029914\times 10^{-7}$&5.6$\times 10^{-10}$\\
0&10&0.7&45$^{\circ}$&$-3.55415027484\times 10^{-7}$&7.4$\times 10^{-9}$\\
0&10&0.7&70$^{\circ}$&$-3.55415029916\times 10^{-7}$&5.6$\times 10^{-10}$\\
  \hline
0&10&0.9& 0$^{\circ}$&$-3.65214284306\times 10^{-7}$& $[10^{-22}]$\\
0&10&0.9&20$^{\circ}$&$-3.65214284171\times 10^{-7}$&3.7$\times 10^{-10}$\\
0&10&0.9&45$^{\circ}$&$-3.65214284814\times 10^{-7}$&1.4$\times 10^{-9}$\\
0&10&0.9&70$^{\circ}$&$-3.65214285085\times 10^{-7}$&2.1$\times 10^{-9}$\\
\hline \hline
\end{tabular}
\label{tab:Schwarz_fluxH}
\end{center}
\end{table}

\begin{figure}[htbp]
\begin{center}
\includegraphics[scale=1.2]{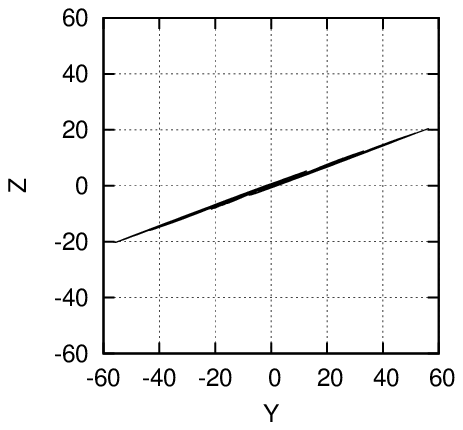}
$\quad$
\includegraphics[scale=1.2]{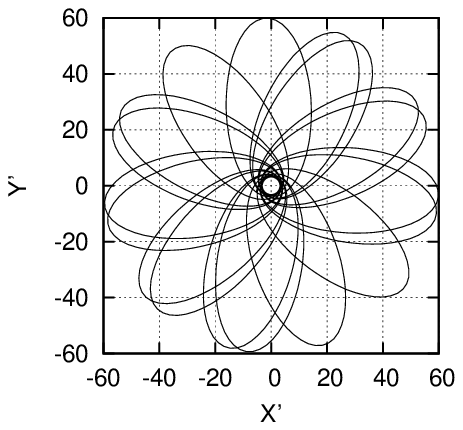}
\end{center}
\caption{Plots of the orbits in the same coordinate systems as those used in
 Fig.~\ref{fig:orbit01}.
This generic geodesic orbit has
eccentricity $e=0.9$, semilatus rectum $p=6M$ and inclination angle
$\theta_{\T{inc}}=20^{\circ}$.
The spin of the black hole is set to $a=0.9M$.
}
\label{fig:orbit0920}
\end{figure}

In Table \ref{tab:dh_table}, we show the rates of change of the
three constants of motion due to the emission of gravitational waves to infinity
in the case of various bound orbits around a Kerr black hole with $a=0.9M$. 
In this table, we use the same orbital parameters as those used by Drasco and
Hughes in Ref.~\citen{Drasco:2006}. 
Our results are consistent with theirs except for the rates of
change of the Carter constant, $\left<d\C{C}/dt\right>$. 
This is because they used formulas for $\left<d\C{C}/dt\right>$ under
the approximation that the inclination angle does not change.
Of course, this holds only approximately. 
This is the first time that the rate of change of the
Carter constant has been computed accurately using an adiabatic approximation.
In Table.~\ref{tab:dh_table}, we also show results for the 
highly eccentricity of $e=0.9$. 
In Fig.~\ref{fig:orbit0920}, we plot a highly eccentric orbit with
$e=0.9$, $p=6M$ and $\theta_{\T{inc}}=20^{\circ}$. 
As indicated in the bottom right figure in Fig.~\ref{fig:nmode2}, the peak 
location of the $n$-mode is approximately $600$ when $\ell=m=20$ and $k=0$.

The values in the square brackets in Table \ref{tab:dh_table} 
are the truncation errors $\Delta_{(\ell_{\rm max})}^\infty$, 
which take values from $10^{-11}$ to $10^{-6}$. 
For highly eccentric cases, $e>0.5$, $\Delta_{(\ell_{\rm max})}$ is larger 
than the truncation error of the $n$- and $k$-mode summation.
Thus, the accuracy for such cases is only limited 
by the truncation of the $\ell$-mode summation.
For $e<0.5$, the truncation errors of the $\ell$-mode summation and $n$- and $k$-mode 
summation are comparable, and both errors contribute to the error of the total flux.

In Table \ref{tab:Kerr_fluxH}, we show the
time-averaged rates of change of the three constants of motion, 
$\left<d\C{E}/dt\right>^{\rm H}$,  $\left<d\C{L}_{z}/dt\right>^{\rm H}$
and $\left<d\C{C}/dt\right>^{\rm H}$, 
due to the absorption at the horizon for
various bound orbits around a Kerr black hole.
The orbital parameters used here are again the same as those used by Drasco
and Hughes in Ref.~\citen{Drasco:2006} except for when $e=0.9$.
Our results are consistent with theirs except for the rate of
change of the Carter constant. 
As in the case of Table \ref{tab:dh_table}, this is the first time that the rate of change of the
Carter constant due to absorption by the black hole 
has been computed accurately using an adiabatic approximation.
The cases when $e=0.9$ are also new results of this work. 
The numbers in square brackets are the truncation errors of the 
$\ell$-mode summation, $\Delta_{(\ell_{\rm max})}$, with $\ell_{\T{max}}=20$.
The truncation error of the $\ell$-mode summation
is much smaller than that of the $n$- and $k$-mode summation,
which is approximately $10^{-10}$. The accuracy of the data in Table \ref{tab:Kerr_fluxH}
is limited by the truncation of the $n$- and $k$-mode summation.

Here, we discuss the sign of $\left<d\C{E}/dt\right>^{\T{H}}$. 
From Table \ref{tab:Kerr_fluxH}, we find that the particle loses energy,
i.e., $\left<d\C{E}/dt\right>^{\T{H}} <0$, 
when $e=0.3, 0.5$ and $0.7$ and $\theta_\T{inc}=80^\circ$, whereas 
the particle gains energy, i.e., $\left<d\C{E}/dt\right>^{\T{H}}>0$,
in the other cases when the eccentricity or the inclination angle is small. 
From Eq.~\eqref{eq:dEdt}, we find that the sign of each mode 
$\left< d\C{E}/dt\right>^\T{H}_{\ell mkn}$ is determined by the sign of $\alpha_{\ell mkn}$,
i.e., $P=\omega_{mkn}-ma/(2Mr_+)$ $=k \Omega_\theta  + n \Omega_r$ $-m (a/(2Mr_+)-\Omega_\phi )$. 
The sign of $\left<d\C{E}/dt\right>^{\T{H}}$ is 
determined by the sign of each $\alpha_{\ell mkn}$ and the absolute value of 
$\left< d\C{E}/dt\right>^\T{H}_{\ell mkn}$.
For $a>0$ (corotation of the particle and the black hole),
when $a$ is large and the modes with $m>0$ and small $k$ and $n$ dominate
the total energy flux, the particle can gain energy. 
We find from Figs.~\ref{fig:nmodeH2} and \ref{fig:kmodeH3} that 
when the eccentricity and inclination angle are small, 
the mode with $\ell=m=2$ and $k=0$ dominates the total energy flux 
and the particle gains energy. 
On the other hand, when the eccentricity and inclination angle are large,
we find from Figs.~\ref{fig:nmodeH2} and \ref{fig:kmodeH4} that 
modes with large $k$ and $n$, which result in $P > 0$, 
contribute to the total energy flux.
Furthermore, Fig.~\ref{fig:kmodeH4} shows that the peak value of 
$m<0$ modes is very similar to that for $m>0$. 
This also contributes to making $\left< d\C{E}/dt\right>^\T{H}$ positive
in the cases of large eccentricity and a large inclination angle.

To confirm the accuracy of the numerical code, we compare our results with
the analytical post-Newtonian formulas for orbits that are slightly
eccentric but highly inclined\cite{Ganz}. 
We show the results for $p=100M$ and various $e$ and $\theta_\T{inc}$
in Table~\ref{tab:PN_compare}.
We find that our numerical results and 
the post-Newtonian formulas agree with an accuracy of $\sim 10^{-4}$ or better.
Note that when $p$ is smaller than $100M$, the accuracy of the post-Newtonian 
formulas becomes worse than this value. 

Once we have the rates of change of the constants of motion $I^i=(\C{E},\C{L}_{z},\C{C})$,
we can derive the rate of change of the orbital elements $\iota^i=(p,e,\theta_\T{inc})$.
We have the following relation, 
\begin{align}
\left\langle\frac{d\iota^i}{dt}\right\rangle = 
\left (G^{-1}\right)^{i}_{j}\left\langle\frac{dI^j}{dt}\right\rangle,
\label{eq:diotadt}
\end{align}
where $G^{i}_{j}=\partial I^i/\partial \iota^j$. 
In Table~\ref{tab:PN_dIdt}, we compare $\left< d\iota^i/dt\right>^\infty$ 
with values derived using the post-Newtonian formulas
by applying Eq.~\eqref{eq:diotadt} to the data in Table~\ref{tab:PN_compare}.
The relative errors of $\left<dp/dt\right>^\infty$ and $\left<de/dt\right>^\infty$ are $10^{-4}$, 
whereas the relative error of $\left< d\theta_\T{inc}/dt\right>^\infty$
is approximately $10^{-2}$. 
This is because in Ref.~\citen{Ganz}, $\left<d\theta_\T{inc}/dt\right>^\infty$ is
derived only up to 1PN order from the leading order, 
whereas $\left<dp/dt\right>$ and $\left<de/dt\right>$ are derived up to 2.5PN order. 

We apply Eq.~\eqref{eq:diotadt} to the data in 
Tables~\ref{tab:dh_table} and ~\ref{tab:Kerr_fluxH} 
to obtain the rates of change of the orbital elements 
due to the emission of gravitational waves to infinity and 
to absorption by the black hole, which are shown in  
Tables~\ref{tab:dIdt8} and ~\ref{tab:dIdtH}, respectively.
We find that in most cases, the flux at infinity and absorption by the black hole
exhibit opposite effects except when 
$e=0.3, 0.5$ and $0.7$ and $\theta_\T{inc}=80^\circ$. 
In all cases, since the flux at infinity dominates the sign of the total rate of change, 
the total rates of change shown in Table \ref{tab:dIdt} have the same sign 
as those for infinity.

\begin{figure}[htbp]
\begin{center}
\includegraphics[scale=1.0]{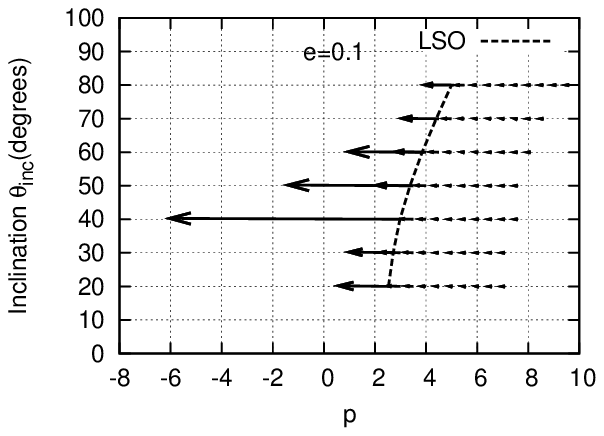}
$\quad$
\includegraphics[scale=1.0]{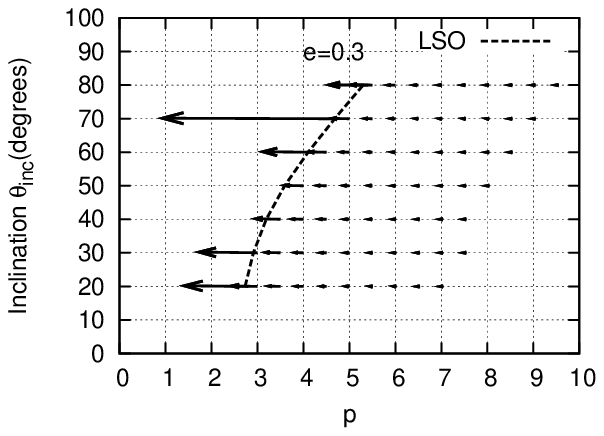}\\
\includegraphics[scale=1.0]{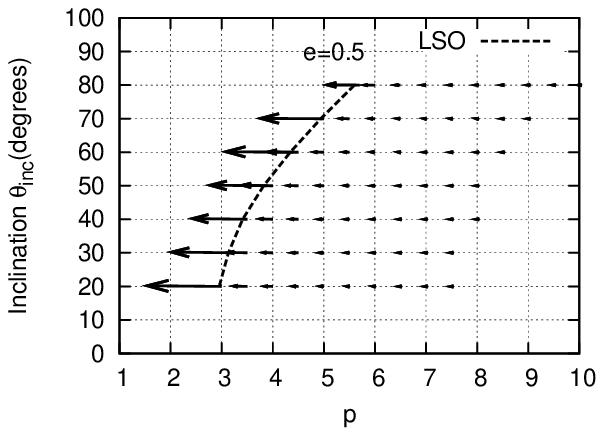}
$\quad$
\includegraphics[scale=1.0]{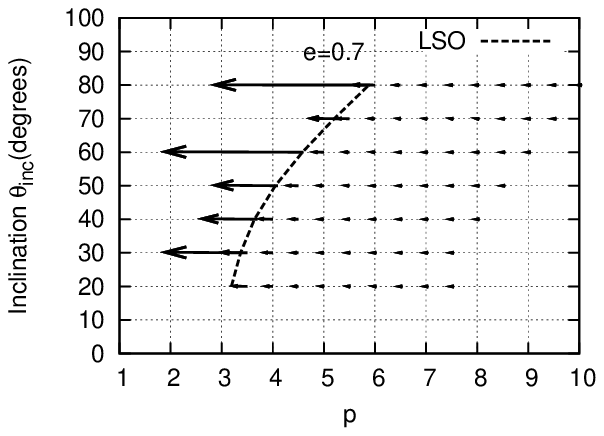}
\caption{
Evolution of eccentric and inclined orbits projected 
on the $e={\T{const.}}$ plane. The black hole spin is $a=0.9M$. 
The dashed curves represent the last stable orbit (LSO). 
Each point represents an orbit and each arrow represents 
the rate of change of the orbit $(\left<dp/dt\right>,\left<d\theta_{\T{inc}}/dt\right>)$.
Here, the lengths of arrows are normalized appropriately. 
The top left and top right figures show the evolution on the $e=0.1$ and $e=0.3$ planes 
respectively. The bottom left and bottom right figures show the evolution on the $e=0.5$ and 
$e=0.7$ planes respectively. 
At a large distance, $p$ is always decreasing and $\theta_{\T{inc}}$ is always 
increasing. In the computation of this figure, we set $\ell_{\T{max}}=5$. 
}
\label{fig:dIdt_theta}
\end{center}
\end{figure}

\begin{figure}[htbp]
\begin{center}
\includegraphics[scale=1.0]{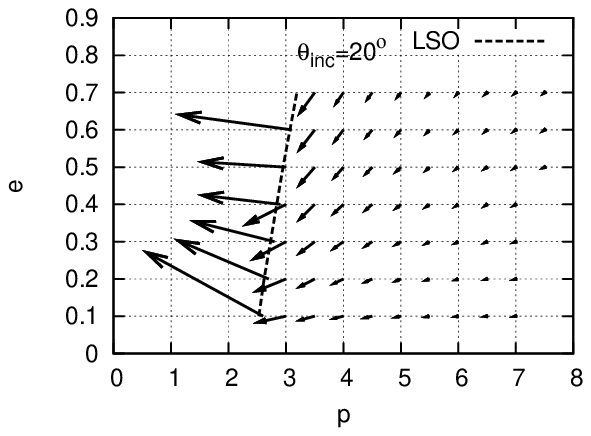}
$\quad$
\includegraphics[scale=1.0]{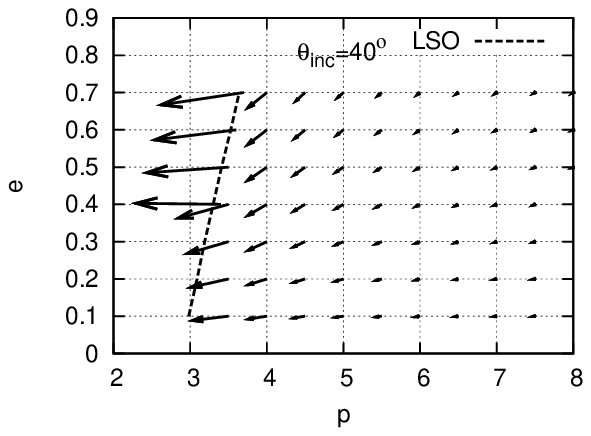}\\
\includegraphics[scale=1.0]{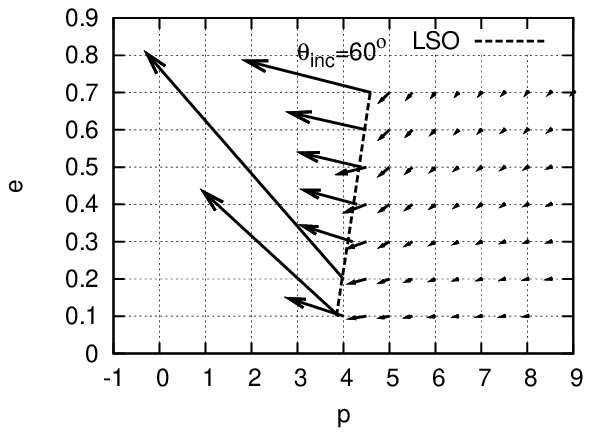}
$\quad$
\includegraphics[scale=1.0]{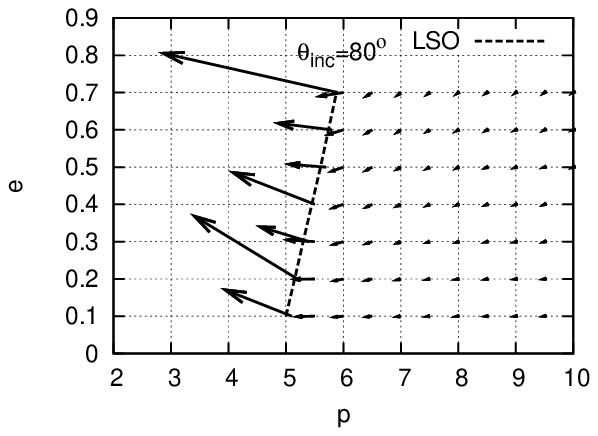}
\caption{
Evolution of eccentric and inclined orbits projected 
on the $\theta_{\T{inc}}={\T{const.}}$ plane. The black hole spin is 
$a=0.9M$. The dashed curves represent the last stable orbit (LSO). 
Each orbit is represented by a point and its evolution is
represented by a vector $(\left<dp/dt\right>,\left<de/dt\right>)$.
The top left figure shows the evolution on the $\theta_{\T{inc}}=20^\circ$ plane 
and the top right figure shows the evolution on the $\theta_{\T{inc}}=40^\circ$ plane.
The bottom left figure shows the evolution on the $\theta_{\T{inc}}=60^\circ$ plane 
and the bottom right figure shows the evolution on the $\theta_{\T{inc}}=80^\circ$ plane. 
At a large distance, $p$ and $e$ are always decreasing.
In the computation of this figure, we set $\ell_{\T{max}}=5$. 
}
\label{fig:dIdt_ecc}
\end{center}
\end{figure}

To demonstrate some aspects of the evolution of the orbital elements, 
in Figs. \ref{fig:dIdt_theta} and \ref{fig:dIdt_ecc}
we plot the total rates of change $(\left<dp/dt\right>, \left<d\theta_{\T{inc}}/dt\right>)$ and
$(\left<dp/dt\right>, \left<de/dt\right>)$ on the $(p,\theta_{\T{inc}})$ and $(p,e)$ planes,
respectively, in the case when $a=0.9M$ and $\ell_{\T{max}}=5$. 
We find that although the eccentricity is always decreasing at large $p$,
it can increase near the last stable orbit (LSO). The change in the inclination angle is not very 
significant in the figures, but it is always increasing at large $p$.

\begin{table}[htbp]
%\begin{center}
\caption{
Time-averaged rates of change of the three constants of motion, energy
$\left<d\C{E}/dt\right>^{\infty}$, 
angular momentum $\left<d\C{L}_{z}/dt\right>^{\infty}$
and the Carter constant $\left<d\C{C}/dt\right>^{\infty}$, 
due to gravitational waves radiated to infinity per unit mass for various
generic orbits around a Kerr black hole.
The orbital parameters used here are the same as those used by Drasco
and Hughes in Ref.~\citen{Drasco:2006} except when $e=0.9$. 
Our results are consistent with theirs except for the rate of
change of the Carter constant, $\left<d\C{C}/dt\right>^{\infty}$.
Numbers in square brackets are the truncation errors of the 
$\ell$-mode summation, $\Delta_{(\ell_{\rm max})}$ with $\ell_{\T{max}}=20$.
Note that the case of $a=0.9M$, $p=6M$, $e=0.9$ and 
$\theta_{\T{inc}}=80^{\circ}$ does not result in
stable bound orbits. 
}
\scalebox{0.8}{
\begin{minipage}{\textwidth}
\begin{tabular}{c c c c || c c c}
\hline \hline
  $a/M$ & $p/M$ & $e$ & $\theta_{\T{inc}}$
& $\left<\left. d\C{E}\right/ dt \right >^{\infty} $
& $\left<\left. d\C{L}_{z} \right/ dt \right >^{\infty} $
& $\left<\left. d\C{C}\right/ dt \right >^{\infty} $\\
\hline
$0.9$ & $6$ & $0.1$ & $20^{\circ}$ &
$\ -5.87363800087\times10^{-4}[10^{-11}]\ $&
$\ -8.53727881580\times10^{-3}\ $& 
$\ -5.24019848546\times10^{-3}\ $\\
$0.9$ & $6$ & $0.1$ & $40^{\circ}$ &
$\ -6.18322941497\times10^{-4}[10^{-11}]\ $ &
$\ -7.63099401313\times10^{-3}\ $ &
$\ -2.02271137874\times10^{-2}\ $\\
$0.9$ & $6$ & $0.1$ & $60^{\circ}$ &
$\ -6.83348195277\times10^{-4}[10^{-11}]\ $ &
$\ -6.07829111698\times10^{-3}\ $ & 
$\ -4.32194650210\times10^{-2}\ $\\
$0.9$ & $6$ & $0.1$ & $80^{\circ}$ &
$\ -8.05858117692\times10^{-4}[10^{-10}]\ $ &
$\ -3.62538058308\times10^{-3}\ $  & 
$\ -7.18520476701\times10^{-2}\ $\\ \hline
$0.9$ & $6$ & $0.3$ & $20^{\circ}$ & 
$\ -6.80409929713\times10^{-4}[10^{-9}]\ $&
$\ -8.62590762129\times10^{-3}\ $  & 
$\ -5.22145052503\times10^{-3}\ $\\
$0.9$ & $6$ & $0.3$ & $40^{\circ}$ &
$\ -7.26541780924\times10^{-4}[10^{-9}]\ $ &
$\ -7.84019287653\times10^{-3}\ $  & 
$\ -2.04389407707\times10^{-2}\ $\\
$0.9$ & $6$ & $0.3$ & $60^{\circ}$ &
$\ -8.30597576584\times10^{-4}[10^{-9}]\ $ &
$\ -6.49674204013\times10^{-3}\ $  & 
$\ -4.50701803710\times10^{-2}\ $\\
$0.9$ & $6$ & $0.3$ & $80^{\circ}$ &
$\ -1.08394107072\times10^{-3}[10^{-9}]\ $  &
$\ -4.38279817141\times10^{-3}\ $ & 
$\ -8.18315782169\times10^{-2}\ $\\ \hline
$0.9$ & $6$ & $0.5$ & $20^{\circ}$ &
$\ -7.98925629079\times10^{-4}[10^{-8}]\ $ &
$\ -8.34750401557\times10^{-3}\ $  & 
$\ -4.94704500000\times10^{-3}\ $\\
$0.9$ & $6$ & $0.5$ & $40^{\circ}$ &
$\ -8.74335722008\times10^{-4}[10^{-8}]\ $  &
$\ -7.81941309824\times10^{-3}\ $  & 
$\ -1.98900691857\times10^{-2}\ $\\
$0.9$ & $6$ & $0.5$ & $60^{\circ}$ &
$\ -1.05884649558\times10^{-3}[10^{-8}]\ $  &
$\ -6.95065613502\times10^{-3}\ $  & 
$\ -4.65662839707\times10^{-2}\ $\\
$0.9$ & $6$ & $0.5$ & $80^{\circ}$ &
$\ -1.67699406035\times10^{-3}[10^{-7}]\ $  &
$\ -5.90867286937\times10^{-3}\ $  & 
$\ -1.01809636563\times10^{-1}\ $\\ \hline
$0.9$ & $6$ & $0.7$ & $20^{\circ}$ &
$\ -7.73126177805\times10^{-4}[10^{-7}]\ $&
$\ -6.69310061924\times10^{-3}\ $& 
$\ -3.88688412163\times10^{-3}\ $\\
$0.9$ & $6$ & $0.7$ & $40^{\circ}$ &
$\ -8.75195550414\times10^{-4}[10^{-7}]\ $  &
$\ -6.53053154749\times10^{-3}\ $  & 
$\ -1.62422254375\times10^{-2}\ $\\
$0.9$ & $6$ & $0.7$ & $60^{\circ}$ &
$\ -1.14691287812\times10^{-3}[10^{-7}]\ $  &
$\ -6.38027895400\times10^{-3}\ $  & 
$\ -4.14632893132\times10^{-2}\ $\\
$0.9$ & $6$ & $0.7$ & $80^{\circ}$ &
$\ -2.71933022663\times10^{-3}[10^{-6}]\ $  &
$\ -8.40183718587\times10^{-3}\ $  & 
$\ -1.34155501090\times10^{-1}\ $\\ \hline
$0.9$ & $6$ & $0.9$ & $20^{\circ}$ &
$\ -3.22243369277\times10^{-4}[10^{-6}]\ $ &
$\ -2.36914164090\times10^{-3}\ $ & 
$\ -1.35698436703\times10^{-3}\ $\\
$0.9$ & $6$ & $0.9$ & $40^{\circ}$ &
$\ -3.81407518944\times10^{-4}[10^{-6}]\ $ &
$\ -2.43273577381\times10^{-3}\ $ & 
$\ -5.96550294295\times10^{-3}\ $\\
$0.9$ & $6$ & $0.9$ & $60^{\circ}$ &
$\ -5.55433712227\times10^{-4}[10^{-6}]\ $ &
$\ -2.68099758625\times10^{-3}\ $  & 
$\ -1.71022011422\times10^{-2}\ $\\
$0.9$ & $6$ & $0.9$ & $80^{\circ}$ &
$\ - \ $ &
$\ - \ $  & 
$\ - \ $\\
\hline \hline
\end{tabular}
\end{minipage}
}
\label{tab:dh_table}
\end{table}

\begin{table}[htbp]
\caption{
Time-averaged rates of change of the three constants of motion, energy
$\left<d\C{E}/dt\right>^{\rm H}$, 
angular momentum $\left<d\C{L}_{z}/dt\right>^{\rm H}$
and the Carter constant $\left<d\C{C}/dt\right>^{\rm H}$
due to gravitational waves absorbed at the horizon per unit mass for various
generic orbits around a Kerr black hole.
The orbital parameters used here are the same as those used by Drasco
and Hughes in Ref.~\citen{Drasco:2006} except when $e=0.9$.
Our results are consistent with theirs except for the rate of
change of the Carter constant, $\left<d\C{C}/dt\right>^{\rm H}$.
Numbers in square brackets are the truncation errors of the 
$\ell$-mode summation, $\Delta_{(\ell_{\rm max})}$ with $\ell_{\T{max}}=20$.
The case of $q=0.9M$, $p=6M$, $e=0.9$ and 
$\theta_{\T{inc}}=80^{\circ}$ does not result in stable bound orbits. 
}
\scalebox{0.78}{
\begin{minipage}{\textwidth}
\begin{tabular}{c c c c || c c c}
\hline \hline
  $\ a/M\ $ & $\ \ p/M\ \ $ & $\ \ {e}\ \ $ & $\theta_{\T{inc}}$
& $\left<\left. d\C{E}\right/ dt \right >^{\rm H} $
& $\left<\left. d\C{L}_{z} \right/ dt \right >^{\rm H} $
& $\left<\left. d\C{C}\right/ dt \right >^{\rm H} $\\
\hline
$0.9$ & $6$ & $0.1$ & $20^{\circ}$ &
$\ 4.25245612585\times 10^{-6} [10^{-25}]\ $&
$\ 6.71500254679\times 10^{-5} \ $& 
$\ 1.41073696418\times 10^{-6} $\\
$0.9$ & $6$ & $0.1$ & $40^{\circ}$ &
$\  3.94882721384\times 10^{-6} [10^{-27}]\ $ &
$\  7.73637887857\times 10^{-5} \ $ &
$\ -2.22131838862\times 10^{-5}\ $\\
$0.9$ & $6$ & $0.1$ & $60^{\circ}$ &
$\  3.33113477148\times 10^{-6} [10^{-30}]\ $ &
$\  1.11677030642\times 10^{-4} \ $ & 
$\ -1.02353953718\times 10^{-4}\ $\\
$0.9$ & $6$ & $0.1$ & $80^{\circ}$ &
$\  9.50601680011\times 10^{-7} [10^{-35}]\ $ &
$\  1.90137812316\times 10^{-4} \ $  & 
$\ -2.72562414552\times 10^{-4}\ $\\ \hline
$0.9$ & $6$ & $0.3$ & $20^{\circ}$ & 
$\  5.86967815445\times 10^{-6} [10^{-23}]\ $&
$\  7.76727457985\times 10^{-5} \ $  & 
$\ -5.86205174377\times 10^{-6}\ $\\
$0.9$ & $6$ & $0.3$ & $40^{\circ}$ &
$\  5.84180692574\times 10^{-6} [10^{-23}]\ $ &
$\  1.00066438717\times 10^{-4} \ $  & 
$\ -5.95214221681\times 10^{-5}\ $\\
$0.9$ & $6$ & $0.3$ & $60^{\circ}$ &
$\  5.19494530315\times 10^{-6} [10^{-26}]\ $ &
$\  1.66116835805\times 10^{-4} \ $  & 
$\ -2.26658646533\times 10^{-4}\ $\\
$0.9$ & $6$ & $0.3$ & $80^{\circ}$ &
$\ -2.95984810527\times 10^{-9}[10^{-22}]\ $  &
$\  3.45153570099\times 10^{-4} \ $ & 
$\ -7.00228568783\times 10^{-4}\ $\\ \hline
$0.9$ & $6$ & $0.5$ & $20^{\circ}$ &
$\  8.34425799664\times 10^{-6} [10^{-21}]\ $ &
$\  9.13622955381\times 10^{-5} \ $  & 
$\ -2.13883155844\times 10^{-5}\ $\\
$0.9$ & $6$ & $0.5$ & $40^{\circ}$ &
$\  8.94532600748\times 10^{-6} [10^{-21}]\ $  &
$\  1.37186174930\times 10^{-4} \ $  & 
$\ -1.42441848989\times 10^{-4}\ $\\
$0.9$ & $6$ & $0.5$ & $60^{\circ}$ &
$\  8.08290137218\times 10^{-6} [10^{-23}]\ $  &
$\  2.69931271416\times 10^{-4} \ $  & 
$\ -5.34259432172\times 10^{-4}\ $\\
$0.9$ & $6$ & $0.5$ & $80^{\circ}$ &
$\ -5.98308999615\times 10^{-6}[10^{-17}]\ $  &
$\  7.40794676040\times 10^{-4} \ $  & 
$\ -2.07799385090\times 10^{-3}\ $\\ \hline
$0.9$ & $6$ & $0.7$ & $20^{\circ}$ &
$\  9.29526284834\times 10^{-6} [10^{-19}]\ $&
$\  8.90473867250\times 10^{-5} \ $& 
$\ -3.96683923123\times 10^{-5}\ $\\
$0.9$ & $6$ & $0.7$ & $40^{\circ}$ &
$\  1.05570527008\times 10^{-5} [10^{-19}]\ $  &
$\  1.56425173930\times 10^{-4} \ $  & 
$\ -2.43302981071\times 10^{-4}\ $\\
$0.9$ & $6$ & $0.7$ & $60^{\circ}$ &
$\  9.20041590887\times 10^{-6} [10^{-21}]\ $  &
$\  3.62722481332\times 10^{-4} \ $  & 
$\ -9.65381244057\times 10^{-4}\ $\\
$0.9$ & $6$ & $0.7$ & $80^{\circ}$ &
$\ -2.74019070298\times 10^{-5}[10^{-13}]\ $  &
$\  1.59338857620\times 10^{-3} \ $  & 
$\ -5.85248852167\times 10^{-3}\ $\\ \hline
$0.9$ & $6$ & $0.9$ & $20^{\circ}$ &
$\  4.17773280477\times 10^{-6} [10^{-18}]\ $ &
$\  3.74778275669\times 10^{-5} \ $ & 
$\ -2.72898298129\times 10^{-5}\ $\\
$0.9$ & $6$ & $0.9$ & $40^{\circ}$ &
$\  4.94483767585\times 10^{-6} [10^{-18}]\ $ &
$\  7.68457866865\times 10^{-5} \ $ & 
$\ -1.65632369297\times 10^{-4}\ $\\
$0.9$ & $6$ & $0.9$ & $60^{\circ}$ &
$\ 3.94007307443\times 10^{-6} [10^{-14}]\ $ &
$\ 2.10536770888\times 10^{-4} \ $  & 
$\ -7.17182301393\times 10^{-4}\ $\\
$0.9$ & $6$ & $0.9$ & $80^{\circ}$ &
$\ - \ $ &
$\ - \ $  & 
$\ - \ $\\
\hline \hline
\end{tabular}
\end{minipage}
}
\label{tab:Kerr_fluxH}
\end{table}

\clearpage
\begin{table}[htbp]
\caption{
Comparison of the time-averaged rates of change of the three constants of motion 
derived using our numerical method and analytical post-Newtonian expressions~\cite{Ganz}
for orbits that are slightly eccentric but greatly inclined 
in the case of $a=0.9M$ and $p=100M$.
Our numerical results are consistent with the post-Newtonian results. 
Relative errors are always approximately $10^{-4}$.
}
\scalebox{0.67}{
\begin{minipage}{\textwidth}
\begin{tabular}{c c || c c c| c c c}
\hline \hline
  $\ \ {e}\ \ $ & $\theta_{\T{inc}}$
& $\left<\left. d\C{E}\right/ dt \right >^{\infty}_{\rm Numerical} $
& $\left<\left. d\C{L}_{z} \right/ dt \right >^{\infty}_{\rm Numerical} $
& $\left<\left. d\C{C}\right/ dt \right >^{\infty}_{\rm Numerical} $
& $\left<\left. d\C{E}\right/ dt \right >^{\infty}_{\rm Post-Newton} $
& $\left<\left. d\C{L}_{z} \right/ dt \right >^{\infty}_{\rm Post-Newton} $
& $\left<\left. d\C{C}\right/ dt \right >^{\infty}_{\rm Post-Newton} $\\
\hline
$0.01$ & $20^{\circ}$ &
$\ -6.2072\times10^{-10}\ $&
$\ -5.8382\times10^{-7}\ $& 
$\ -1.4687\times10^{-6}\ $&
$\ -6.2067\times10^{-10}\ $&
$\ -5.8376\times10^{-7}\ $& 
$\ -1.4687\times10^{-6}\ $\\
$0.01$ & $45^{\circ}$ &
$\ -6.2142\times10^{-10}\ $&
$\ -4.4029\times10^{-7}\ $& 
$\ -6.2905\times10^{-6}\ $&
$\ -6.2136\times10^{-10}\ $ &
$\ -4.4024\times10^{-7}\ $ &
$\ -6.2901\times10^{-6}\ $\\
$0.01$ & $70^{\circ}$ &
$\ -6.2259\times10^{-10} $&
$\ -2.1420\times10^{-7}\ $& 
$\ -1.1146\times10^{-5}\ $&
$\ -6.2251\times10^{-10}\ $ &
$\ -2.1415\times10^{-7}\ $ & 
$\ -1.1145\times10^{-5}\ $\\\hline
$0.05$ & $20^{\circ}$ &
$\ -6.2302\times10^{-10}\ $&
$\ -5.8299\times10^{-7}\ $& 
$\ -1.4667\times10^{-6}\ $&
$\ -6.2296\times10^{-10}\ $&
$\ -5.8293\times10^{-7}\ $& 
$\ -1.4666\times10^{-6}\ $\\
$0.05$ & $45^{\circ}$ &
$\ -6.2373\times10^{-10}\ $&
$\ -4.3967\times10^{-7}\ $& 
$\ -6.2816\times10^{-6}\ $&
$\ -6.2366\times10^{-10}\ $ &
$\ -4.3962\times10^{-7}\ $ &
$\ -6.2812\times10^{-6}\ $\\
$0.05$ & $70^{\circ}$ &
$\ -6.2490\times10^{-10}\ $&
$\ -2.1390\times10^{-7}\ $& 
$\ -1.1130\times10^{-5}\ $&
$\ -6.2481\times10^{-10}\ $ &
$\ -2.1386\times10^{-7}\ $ & 
$\ -1.1129\times10^{-5}\ $\\\hline
$0.09$ & $20^{\circ}$ &
$\ -6.2827\times10^{-10}\ $&
$\ -5.8103\times10^{-7}\ $& 
$\ -1.4617\times10^{-6}\ $&
$\ -6.2818\times10^{-10}\ $&
$\ -5.8097\times10^{-7}\ $& 
$\ -1.4616\times10^{-6}\ $\\
$0.09$ & $45^{\circ}$ &
$\ -6.2899\times10^{-10}\ $&
$\ -4.3820\times10^{-7}\ $& 
$\ -6.2606\times10^{-6}\ $&
$\ -6.2890\times10^{-10}\ $ &
$\ -4.3814\times10^{-7}\ $ &
$\ -6.2601\times10^{-6}\ $\\
$0.09$ & $70^{\circ}$ &
$\ -6.3019\times10^{-10} $&
$\ -2.1320\times10^{-7}\ $& 
$\ -1.1093\times10^{-5}\ $&
$\ -6.3008\times10^{-10}\ $ &
$\ -2.1315\times10^{-7}\ $ & 
$\ -1.1092\times10^{-5}\ $\\
\hline \hline
\end{tabular}
\end{minipage}
}
\label{tab:PN_compare}
\end{table}

\begin{table}[htbp]
\caption{
Comparison of the time-averaged rates of change of orbital elements 
derived using our numerical method and the analytical 
post-Newtonian expressions~\cite{Ganz} for 
orbits that are slightly eccentric but greatly inclined
in the case of $a=0.9M$ and $p=100M$.
Our numerical results are consistent with the post-Newtonian results. 
The relative errors of both $\left<dp/dt\right>$ and $\left<de/dt\right>$ 
are always approximately $10^{-4}$. 
However, the relative errors of $\left<d\theta_{\T{inc}}/dt\right>$ are approximately $10^{-2}$ since 
$\left<d\theta_{\T{inc}}/dt\right>^\infty_{\rm Post-Newton}$ in Ref.~\citen{Ganz} is 1PN.
}
\scalebox{0.67}{
\begin{minipage}{\textwidth}
\begin{tabular}{c c || c c c| c c c}
\hline \hline
  $\ \ {e}\ \ $ & $\theta_{\T{inc}}$
& $\left<\left. dp\right/ dt \right >^{\infty}_{\rm Numerical} $
& $\left<\left. de\right/ dt \right >^{\infty}_{\rm Numerical} $
& $\left<\left. d\theta_{\T{inc}}\right/ dt \right >^{\infty}_{\rm Numerical} $
& $\left<\left. dp\right/ dt \right >^{\infty}_{\rm Post-Newton} $
& $\left<\left. de\right/ dt \right >^{\infty}_{\rm Post-Newton} $
& $\left<\left. d\theta_{\T{inc}}\right/ dt \right >^{\infty}_{\rm Post-Newton} $\\
\hline
$0.01$ & $20^{\circ}$ &
$\ -1.2560\times 10^{-5}\ $&
$\ -1.9807\times 10^{-9}\ $& 
$\ 2.5353\times 10^{-9}\ $&
$\ -1.2558\times 10^{-5}\ $&
$\ -1.9803\times 10^{-9}\ $& 
$\ 2.6807\times 10^{-9}\ $\\
$0.01$ & $45^{\circ}$ &
$\ -1.2587\times 10^{-5}\ $&
$\ -1.9851\times 10^{-9}\ $& 
$\ 5.3432\times 10^{-9}\ $&
$\ -1.2585\times 10^{-5}\ $ &
$\ -1.9846\times 10^{-9}\ $ &
$\ 5.5488\times 10^{-9}\ $\\
$0.01$ & $70^{\circ}$ &
$\ -1.2631\times 10^{-5}\ $&
$\ -1.9923\times 10^{-9}\ $& 
$\ 7.3134\times 10^{-9}\ $&
$\ -1.2630\times 10^{-5}\ $ &
$\ -1.9917\times 10^{-9}\ $ & 
$\ 7.3877\times 10^{-9}\ $\\\hline
$0.05$ & $20^{\circ}$ &
$\ -1.2541\times 10^{-5}\ $&
$\ -9.8778\times 10^{-9}\ $& 
$\ 2.5450\times 10^{-9}\ $&
$\ -1.2540\times 10^{-5}\ $&
$\ -9.8753\times 10^{-9}\ $& 
$\ 2.6900\times 10^{-9}\ $\\
$0.05$ & $45^{\circ}$ &
$\ -1.2568\times 10^{-5}\ $&
$\ -9.8996\times 10^{-9}\ $& 
$\ 5.3636\times 10^{-9}\ $&
$\ -1.2567\times 10^{-5}\ $ &
$\ -9.8968\times 10^{-9}\ $ &
$\ 5.5683\times 10^{-9}\ $\\
$0.05$ & $70^{\circ}$ &
$\ -1.2613\times 10^{-5}\ $&
$\ -9.9357\times 10^{-9}\ $& 
$\ 7.3412\times 10^{-9}\ $&
$\ -1.2611\times 10^{-5}\ $ &
$\ -9.9324\times 10^{-9}\ $ & 
$\ 7.4142\times 10^{-9}\ $\\\hline
$0.09$ & $20^{\circ}$ &
$\ -1.2497\times 10^{-5}\ $&
$\ -1.7671\times 10^{-8}\ $& 
$\ 2.5672\times 10^{-9}\ $&
$\ -1.2496\times 10^{-5}\ $&
$\ -1.7666\times 10^{-8}\ $& 
$\ 2.7112\times 10^{-9}\ $\\
$0.09$ & $45^{\circ}$ &
$\ -1.2524\times 10^{-5}\ $&
$\ -1.7711\times 10^{-8}\ $& 
$\ 5.4104\times 10^{-9}\ $&
$\ -1.2523\times 10^{-5}\ $ &
$\ -1.7705\times 10^{-8}\ $ &
$\ 5.6127\times 10^{-9}\ $\\
$0.09$ & $70^{\circ}$ &
$\ -1.2569\times 10^{-5}\ $&
$\ -1.7775\times 10^{-8}\ $& 
$\ 7.4050\times 10^{-9}\ $&
$\ -1.2567\times 10^{-5}\ $ &
$\ -1.7768\times 10^{-8}\ $ & 
$\ 7.4748\times 10^{-9}\ $\\
\hline \hline
\end{tabular}
\end{minipage}}
\label{tab:PN_dIdt}
\end{table}

\begin{table}[htbp]
\caption{
Time-averaged rates of change of orbital elements, 
semilatus rectum $\left<dp/dt\right>^{\infty}$, 
eccentricity $\left<de/dt\right>^{\infty}$
and inclination angle $\left<d\theta_{\T{inc}}/dt\right>^{\infty}$
due to gravitational waves radiated to infinity per unit mass 
for various generic orbits around a Kerr black hole.
The orbital parameters used here are the same as those used 
in Tables \ref{tab:dh_table} and \ref{tab:Kerr_fluxH}. 
Here we set $\ell_{\T{max}}=20$.
Note that the case of $q=0.9M$, $p=6M$, $e=0.9$ and 
$\theta_{\T{inc}}=80^{\circ}$ 
does not result in stable bound orbits. 
}
\scalebox{0.8}{
\begin{minipage}{\textwidth}
\begin{tabular}{c c c c || c c c}
\hline \hline
  $\ a/M\ $ & $\ \ p/M\ \ $ & $\ \ {e}\ \ $ & $\theta_{\T{inc}}$
& $\left< dp/ dt \right >^{\infty} $
& $\left< de/ dt \right >^{\infty} $
& $\left< d\theta_{\T{inc}}/ dt \right >^{\infty} $\\
\hline
$0.9$ & $6$ & $0.1$ & $20^{\circ}$ &
$\ -4.92615496293\times 10^{-2}\ $&
$\ -1.34079103976\times 10^{-3}\ $& 
$\  6.38027263441\times 10^{-3}\ $\\
$0.9$ & $6$ & $0.1$ & $40^{\circ}$ &
$\ -5.67137189535\times 10^{-2}\ $ &
$\ -1.51600653064\times 10^{-3}\ $ &
$\  1.31392492301\times 10^{-2}\ $\\
$0.9$ & $6$ & $0.1$ & $60^{\circ}$ &
$\ -7.66417737679\times 10^{-2}\ $ &
$\ -1.91441435872\times 10^{-3}\ $ & 
$\  2.04198174044\times 10^{-2}\ $\\
$0.9$ & $6$ & $0.1$ & $80^{\circ}$ &
$\ -1.47331549001\times 10^{-1}\ $ &
$\ -2.42219567710\times 10^{-3}\ $  & 
$\  2.80031935585\times 10^{-2}\ $\\ \hline
$0.9$ & $6$ & $0.3$ & $20^{\circ}$ & 
$\ -4.81040940278\times 10^{-2}\ $&
$\ -3.79892010960\times 10^{-3}\ $  & 
$\  7.24540536891\times 10^{-3}\ $\\
$0.9$ & $6$ & $0.3$ & $40^{\circ}$ &
$\ -5.56575271585\times 10^{-2}\ $ &
$\ -4.31276746863\times 10^{-3}\ $  & 
$\  1.50642346054\times 10^{-2}\ $\\
$0.9$ & $6$ & $0.3$ & $60^{\circ}$ &
$\ -7.63519232233\times 10^{-2}\ $ &
$\ -5.50577812857\times 10^{-3}\ $  & 
$\  2.40542698304\times 10^{-2}\ $\\
$0.9$ & $6$ & $0.3$ & $80^{\circ}$ &
$\ -1.62064124200\times 10^{-1}\ $  &
$\ -7.12327756470\times 10^{-3}\ $ & 
$\  3.62128496646\times 10^{-2}\ $\\ \hline
$0.9$ & $6$ & $0.5$ & $20^{\circ}$ &
$\ -4.38342833363\times 10^{-2}\ $ &
$\ -5.44145471464\times 10^{-3}\ $  & 
$\  8.22603779275\times 10^{-3}\ $\\
$0.9$ & $6$ & $0.5$ & $40^{\circ}$ &
$\ -5.12890638716\times 10^{-2}\ $  &
$\ -6.23429803953\times 10^{-3}\ $  & 
$\  1.73940885907\times 10^{-2}\ $\\
$0.9$ & $6$ & $0.5$ & $60^{\circ}$ &
$\ -7.27941856988\times 10^{-2}\ $  &
$\ -8.15895492769\times 10^{-3}\ $  & 
$\  2.91106743461\times 10^{-2}\ $\\
$0.9$ & $6$ & $0.5$ & $80^{\circ}$ &
$\ -2.03777601173\times 10^{-1}\ $  &
$\ -1.10417083320\times 10^{-2}\ $  & 
$\  5.25692497357\times 10^{-2}\ $\\ \hline
$0.9$ & $6$ & $0.7$ & $20^{\circ}$ &
$\ -3.25525935090\times 10^{-2}\ $&
$\ -5.26392055312\times 10^{-3}\ $& 
$\  7.61630914097\times 10^{-3}\ $\\
$0.9$ & $6$ & $0.7$ & $40^{\circ}$ &
$\ -3.88453224067\times 10^{-2}\ $  &
$\ -6.12988746174\times 10^{-3}\ $  & 
$\  1.65083528245\times 10^{-2}\ $\\
$0.9$ & $6$ & $0.7$ & $60^{\circ}$ &
$\ -5.85795749152\times 10^{-2}\ $  &
$\ -8.39195928946\times 10^{-3}\ $  & 
$\  2.95462801593\times 10^{-2}\ $\\
$0.9$ & $6$ & $0.7$ & $80^{\circ}$ &
$\ -4.18519845587\times 10^{-1}\ $  &
$\ -9.50930987412\times 10^{-3}\ $  & 
$\  7.87322089769\times 10^{-2}\ $\\ \hline
$0.9$ & $6$ & $0.9$ & $20^{\circ}$ &
$\ -1.05585939685\times 10^{-2}\ $ &
$\ -2.03522048694\times 10^{-3}\ $ & 
$\  3.00710272834\times 10^{-3}\ $\\
$0.9$ & $6$ & $0.9$ & $40^{\circ}$ &
$\ -1.29931288230\times 10^{-2}\ $ &
$\ -2.43222864060\times 10^{-3}\ $ & 
$\  6.75378303868\times 10^{-3}\ $\\
$0.9$ & $6$ & $0.9$ & $60^{\circ}$ &
$\ -2.16140193063\times 10^{-2}\ $ &
$\ -3.58800678580\times 10^{-3}\ $  & 
$\  1.32815131687\times 10^{-2}\ $\\
$0.9$ & $6$ & $0.9$ & $80^{\circ}$ &
$\ - \ $ &
$\ - \ $  & 
$\ - \ $\\
\hline \hline
\end{tabular}
\end{minipage}}
\label{tab:dIdt8}
\end{table}

\begin{table}[htbp]
\caption{
Time-averaged rates of change of orbital elements, 
semilatus rectum $\left<dp/dt\right>^{\rm H}$, 
eccentricity $\left<de/dt\right>^{\rm H}$
and inclination angle $\left<d\theta_{\T{inc}}/dt\right>^{\rm H}$
due to absorption by a black hole per unit mass for various 
generic orbits around a Kerr black hole.
The orbital parameters used here are the same as those used 
in Tables \ref{tab:dh_table} and \ref{tab:Kerr_fluxH}. 
Here we set $\ell_{\T{max}}=20$.
Note that the case of $q=0.9M$, $p=6M$, $e=0.9$ and 
$\theta_{\T{inc}}=80^{\circ}$ does not result in stable bound
orbits. $\left<dp/dt\right>^{\rm H}$ and $\left<d\theta_{\T{inc}}/dt\right>^{\rm H}$
have opposite signs to 
$\left<dp/dt\right>^{\infty}$ and $\left<d\theta_{\T{inc}}/dt\right>^{\infty}$ respectively. 
$\left<de/dt\right>^{\rm H}$ also has the opposite sign 
to $\left<de/dt\right>^{\infty}$ except when $\theta_\T{inc}=80^\circ$. 
}
\label{tab:dIdtH}
\scalebox{0.8}{
\begin{minipage}{\textwidth}
\begin{tabular}{c c c c || c c c}
\hline \hline
  $\ a/M\ $ & $\ \ p/M\ \ $ & $\ \ {e}\ \ $ & $\theta_{\T{inc}}$
& $\left< dp/ dt \right >^{\rm H} $
& $\left< de/ dt \right >^{\rm H} $
& $\left< d\theta_{\T{inc}}/ dt \right >^{\rm H} $\\
\hline
$0.9$ & $6$ & $0.1$ & $20^{\circ}$ &
$\  3.53847914564\times 10^{-4} \ $&
$\  1.38305197069\times 10^{-5} \ $& 
$\ -4.39326964701\times 10^{-4}\ $\\
$0.9$ & $6$ & $0.1$ & $40^{\circ}$ &
$\  3.69392402044\times 10^{-4} \ $ &
$\  1.43986458994\times 10^{-5} \ $ &
$\ -1.05418631870\times 10^{-3}\ $\\
$0.9$ & $6$ & $0.1$ & $60^{\circ}$ &
$\  4.19452632306\times 10^{-4} \ $ &
$\  1.42953366943\times 10^{-5} \ $ & 
$\ -1.99330099923\times 10^{-3}\ $\\
$0.9$ & $6$ & $0.1$ & $80^{\circ}$ &
$\  4.12137975459\times 10^{-4} \ $ &
$\ -6.18690813899\times 10^{-7}\ $  & 
$\ -3.43092639270\times 10^{-3}\ $\\ \hline
$0.9$ & $6$ & $0.3$ & $20^{\circ}$ & 
$\  3.85261762144\times 10^{-4} \ $&
$\  4.08720376638\times 10^{-5} \ $  & 
$\ -5.76984788510\times 10^{-4}\ $\\
$0.9$ & $6$ & $0.3$ & $40^{\circ}$ &
$\  4.28209164063\times 10^{-4} \ $ &
$\  4.34431544233\times 10^{-5} \ $  & 
$\ -1.47632350406\times 10^{-3}\ $\\
$0.9$ & $6$ & $0.3$ & $60^{\circ}$ &
$\  5.19566999464\times 10^{-4} \ $ &
$\  4.14791122346\times 10^{-5} \ $  & 
$\ -3.07222279960\times 10^{-3}\ $\\
$0.9$ & $6$ & $0.3$ & $80^{\circ}$ &
$\  5.32203519712\times 10^{-4} \ $  &
$\ -2.84117407741\times 10^{-5}\ $ & 
$\ -6.26504606858\times 10^{-3}\ $\\ \hline
$0.9$ & $6$ & $0.5$ & $20^{\circ}$ &
$\  4.13740806210\times 10^{-4} \ $ &
$\  6.26750356393\times 10^{-5} \ $  & 
$\ -8.07328385159\times 10^{-4}\ $\\
$0.9$ & $6$ & $0.5$ & $40^{\circ}$ &
$\  5.05787074091\times 10^{-4} \ $  &
$\  6.85944328970\times 10^{-5} \ $  & 
$\ -2.23246669204\times 10^{-3}\ $\\
$0.9$ & $6$ & $0.5$ & $60^{\circ}$ &
$\  6.80058114128\times 10^{-4} \ $  &
$\  5.95118756580\times 10^{-5} \ $  & 
$\ -5.19405624478\times 10^{-3}\ $\\
$0.9$ & $6$ & $0.5$ & $80^{\circ}$ &
$\  1.05614774875\times 10^{-3} \ $  &
$\ -1.51107876111\times 10^{-4}\ $  & 
$\ -1.34386963554\times 10^{-2}\ $\\ \hline
$0.9$ & $6$ & $0.7$ & $20^{\circ}$ &
$\  3.65047533496\times 10^{-4} \ $&
$\  6.50278285376\times 10^{-5} \ $& 
$\ -9.45012834853\times 10^{-4}\ $\\
$0.9$ & $6$ & $0.7$ & $40^{\circ}$ &
$\  4.98432568124\times 10^{-4} \ $  &
$\  7.32237600001\times 10^{-5} \ $  & 
$\ -2.79399239650\times 10^{-3}\ $\\
$0.9$ & $6$ & $0.7$ & $60^{\circ}$ &
$\  7.75909742923\times 10^{-4} \ $  &
$\  5.50922568705\times 10^{-5} \ $  & 
$\ -7.20072251550\times 10^{-3}\ $\\
$0.9$ & $6$ & $0.7$ & $80^{\circ}$ &
$\  5.63323685413\times 10^{-3} \ $  &
$\ -5.58025680303\times 10^{-4}\ $  & 
$\ -2.85755344237\times 10^{-2}\ $\\ \hline
$0.9$ & $6$ & $0.9$ & $20^{\circ}$ &
$\  1.39603881133\times 10^{-4} \ $ &
$\  2.63466261850\times 10^{-5} \ $ & 
$\ -4.80569511908\times 10^{-4}\ $\\
$0.9$ & $6$ & $0.9$ & $40^{\circ}$ &
$\  2.16553248880\times 10^{-4} \ $ &
$\  3.07560328396\times 10^{-5} \ $ & 
$\ -1.49442340915\times 10^{-3}\ $\\
$0.9$ & $6$ & $0.9$ & $60^{\circ}$ &
$\ 4.19802911304\times 10^{-4} \ $ &
$\ 2.12027539090\times 10^{-5} \ $  & 
$\ -4.26041701312\times 10^{-3}\ $\\
$0.9$ & $6$ & $0.9$ & $80^{\circ}$ &
$\ - \ $ &
$\ - \ $  & 
$\ - \ $\\
\hline \hline
\end{tabular}
\end{minipage}}
\end{table}

\begin{table}[htbp]
\caption{
Total time-averaged rates of change of orbital elements, 
semilatus rectum $\left<dp/dt\right>$, 
eccentricity $\left<de/dt\right>$
and inclination angle $\left<d\theta_{\T{inc}}/dt\right>$
due to gravitational waves per unit mass for various
generic orbits around a Kerr black hole.
The values in this table are the sums of the values 
in Tables \ref{tab:dIdt8} and \ref{tab:dIdtH}.
}
\label{tab:dIdt}
\scalebox{0.8}{
\begin{minipage}{\textwidth}
\begin{tabular}{c c c c || c c c}
\hline \hline
  $\ a/M\ $ & $\ \ p/M\ \ $ & $\ \ {e}\ \ $ & $\theta_{\T{inc}}$
& $\left< dp/ dt \right > $
& $\left< de/ dt \right > $
& $\left< d\theta_{\T{inc}}/ dt \right > $\\
\hline
$0.9$ & $6$ & $0.1$ & $20^{\circ}$ &
$\ -4.89077017147\times 10^{-2}\ $&
$\ -1.32696052005\times 10^{-3}\ $& 
$\  5.94094566971\times 10^{-3}\ $\\
$0.9$ & $6$ & $0.1$ & $40^{\circ}$ &
$\ -5.63443265515\times 10^{-2}\ $ &
$\ -1.50160788474\times 10^{-3}\ $ &
$\  1.20850629114\times 10^{-2}\ $\\
$0.9$ & $6$ & $0.1$ & $60^{\circ}$ &
$\ -7.62223211356\times 10^{-2}\ $ &
$\ -1.90011902203\times 10^{-3}\ $ & 
$\  1.84265164052\times 10^{-2}\ $\\
$0.9$ & $6$ & $0.1$ & $80^{\circ}$ &
$\ -1.46919411025\times 10^{-1}\ $ &
$\ -2.42281436791\times 10^{-3}\ $  & 
$\  2.45722671658\times 10^{-2}\ $\\ \hline
$0.9$ & $6$ & $0.3$ & $20^{\circ}$ & 
$\ -4.77188322657\times 10^{-2}\ $&
$\ -3.75804807194\times 10^{-3}\ $  & 
$\  6.66842058040\times 10^{-3}\ $\\
$0.9$ & $6$ & $0.3$ & $40^{\circ}$ &
$\ -5.52293179945\times 10^{-2}\ $ &
$\ -4.26932431421\times 10^{-3}\ $  & 
$\  1.35879111013\times 10^{-2}\ $\\
$0.9$ & $6$ & $0.3$ & $60^{\circ}$ &
$\ -7.58323562239\times 10^{-2}\ $ &
$\ -5.46429901633\times 10^{-3}\ $  & 
$\  2.09820470308\times 10^{-2}\ $\\
$0.9$ & $6$ & $0.3$ & $80^{\circ}$ &
$\ -1.61531920680\times 10^{-1}\ $  &
$\ -7.15168930548\times 10^{-3}\ $ & 
$\  2.99478035960\times 10^{-2}\ $\\ \hline
$0.9$ & $6$ & $0.5$ & $20^{\circ}$ &
$\ -4.34205425300\times 10^{-2}\ $ &
$\ -5.37877967900\times 10^{-3}\ $  & 
$\  7.41870940759\times 10^{-3}\ $\\
$0.9$ & $6$ & $0.5$ & $40^{\circ}$ &
$\ -5.07832767975\times 10^{-2}\ $  &
$\ -6.16570360664\times 10^{-3}\ $  & 
$\  1.51616218986\times 10^{-2}\ $\\
$0.9$ & $6$ & $0.5$ & $60^{\circ}$ &
$\ -7.21141275847\times 10^{-2}\ $  &
$\ -8.09944305203\times 10^{-3}\ $  & 
$\  2.39166181013\times 10^{-2}\ $\\
$0.9$ & $6$ & $0.5$ & $80^{\circ}$ &
$\ -2.02721453424\times 10^{-1}\ $  &
$\ -1.11928162082\times 10^{-2}\ $  & 
$\  3.91305533802\times 10^{-2}\ $\\ \hline
$0.9$ & $6$ & $0.7$ & $20^{\circ}$ &
$\ -3.21875459755\times 10^{-2}\ $&
$\ -5.19889272458\times 10^{-3}\ $& 
$\  6.67129630612\times 10^{-3}\ $\\
$0.9$ & $6$ & $0.7$ & $40^{\circ}$ &
$\ -3.83468898386\times 10^{-2}\ $  &
$\ -6.05666370174\times 10^{-3}\ $  & 
$\  1.37143604280\times 10^{-2}\ $\\
$0.9$ & $6$ & $0.7$ & $60^{\circ}$ &
$\ -5.78036651723\times 10^{-2}\ $  &
$\ -8.33686703259\times 10^{-3}\ $  & 
$\  2.23455576438\times 10^{-2}\ $\\
$0.9$ & $6$ & $0.7$ & $80^{\circ}$ &
$\ -4.12886608733\times 10^{-1}\ $  &
$\ -1.00673355544\times 10^{-2}\ $  & 
$\  5.01566745532\times 10^{-2}\ $\\ \hline
$0.9$ & $6$ & $0.9$ & $20^{\circ}$ &
$\ -1.04189900874\times 10^{-2}\ $ &
$\ -2.00887386075\times 10^{-3}\ $ & 
$\  2.52653321643\times 10^{-3}\ $\\
$0.9$ & $6$ & $0.9$ & $40^{\circ}$ &
$\ -1.27765755741\times 10^{-2}\ $ &
$\ -2.40147260776\times 10^{-3}\ $ & 
$\  5.25935962953\times 10^{-3}\ $\\
$0.9$ & $6$ & $0.9$ & $60^{\circ}$ &
$\ -2.11942163950\times 10^{-2}\ $ &
$\ -3.56680403189\times 10^{-3}\ $  & 
$\  9.02109615557\times 10^{-3}\ $\\
$0.9$ & $6$ & $0.9$ & $80^{\circ}$ &
$\ - \ $ &
$\ - \ $  & 
$\ - \ $\\
\hline \hline
\end{tabular}
\end{minipage}}
\end{table}

%%%%%%%%%%%%%%%%%%%%%%%%%%%%%%%%%%%%%%%%%%
\section{Summary}\label{sec:summary}
%%%%%%%%%%%%%%%%%%%%%%%%%%%%%%%%%%%%%%%%%%

In this paper, we developed a numerical code to compute 
gravitational waves induced by a particle orbiting around a Kerr black hole. 
We obtained the rates of change of energy, angular momentum and the Carter 
constant for various eccentric and inclined orbits. 
This is the first time that the rate of change of the Carter constant
has been evaluated accurately. 
These computations include highly eccentric cases,  i.e., $e=0.9$.
In previous works, such high eccentricity was not treated. 

Our numerical method is mainly divided into four parts:
the computation of the radial and polar motion, 
the homogeneous solution of the Teukolsky equation, 
the integration of Eq.~\eqref{eq:zlmkn} and the mode summation 
in Eqs. (\ref{eq:dEdt})--(\ref{eq:dIdt}). 
We found that the radial and polar motion can be expressed 
in terms of Jacobi elliptic functions. 
This enabled us to solve the geodesic motion
more accurately than the method in which the equations of motion 
are integrated numerically. 
The asymptotic amplitudes, such as $B_{\ell m\omega}^{\rm inc}$, can be 
computed directly from the MST formalism. 
We do not need to evaluate the homogeneous solutions at a very large distance
to obtain the asymptotic amplitudes. 
We computed the homogeneous Teukolsky solution, $R_{\ell m\omega}^{\rm in}$, 
at a radius $r$  
between $r_{\rm min}\le r\le r_{\rm max}$ using the MST formalism accurately. 
The homogeneous solution at other radial points 
is computed by successive Taylor series expansions. 
The Taylor series method gives very high accuracy and 
is much faster than the use of the hypergeometric function expansion at all radial points.
We computed the integral, Eq.~\eqref{eq:zlmkn}, using the trapezium rule,
which gives very high accuracy when periodic 
functions are integrated over one period.

The accuracy of the numerical results are limited by the truncation of 
the summation of the $\ell$-, $k$- and $n$-modes.
We have verified the behavior of the energy spectrum of the $k$- and $n$-modes. 
We determined the range of the summation of $k$ and $n$ 
to obtain an error due to the truncation of $k$- and $n$-modes of less than $10^{-10}$.
We truncated the $\ell$-mode at $\ell=20$. This value was chosen  
to reduce computation time. 
The error due to the truncation of the $\ell$-mode depends on the orbital 
parameters. When the eccentricity is small, i.e. $e<0.3$, this error is approximately $10^{-9}$.
However, when the eccentricity is $e=0.9$, this error becomes $10^{-5}$,
which is the largest error out of the results computed in this paper. 
Note that since the error is only limited by the truncation of the $\ell$-, $k$- and $n$-modes,
it is straightforward to improve the accuracy. 

To confirm the accuracy of our code, 
we computed the energy flux from a Schwarzschild black hole 
for cases when the orbits are inclined with respect to the equatorial plane.
In the Schwarzschild case, the energy flux should not depend on the 
inclination angle. 
Thus, we can estimate the accuracy of the code by comparing 
the results for inclined orbits and equatorial orbits.
We found that the accuracy of our code is consistent with 
the estimates of the truncation errors of the $\ell$-, $k$- and $n$-modes. 

Although we have not completely optimized the code, 
we briefly discuss the computation time here. 
In the current code, the computation time for one mode is roughly $0.1-0.3$ seconds.
The total computation time is about $6-12$ hours
when $q=0.9, p=6M, e=0.1$ and $\theta_{\rm inc}=20^\circ-80^\circ$,
about 1 day when $q=0.9, p=6M, e=0.7$ and $\theta_{\rm inc}=20^\circ-80^\circ$,
and $1-3$ days when $q=0.9, p=6M, e=0.9$ and $\theta_{\rm inc}=20^\circ-60^\circ$.
These times are for the results of computation with one 2.5 GHz AMD Opteron CPU.
In Fig.1 in Ref. \citen{Hughes:2005qb}, 
the computation time for one mode is shown. 
The computation time using 8 CPUs is about 0.4 seconds 
when $q=0.7, p=10M, e=0.5$ and $\theta_{\rm inc}=45^\circ$.
Thus, the computation time using one CPU will be a few seconds.
Thus, our computation time appears to be much shorter than
that in Ref.\citen{Hughes:2005qb}.

By optimizing the code, we can further increase its speed. 
Also, it is possible to consider variants of the method 
used to obtain the homogeneous solutions, the orbital motion, and so forth.
Although we have not used numerical integration methods, 
it may be advantageous to use them under some circumstances. 
We will consider these issues in the future. 
We will also investigate the computation of gravitational waves including 
the effects of the adiabatic evolution of a particle orbit due to the emission of
gravitational waves. This is important for investigating EMRI
through the data analysis of LISA. 
We will also investigate the possibility of computation over a
wider range of  orbital parameters, such as an eccentricity larger than $0.9$, 
and an inclination angle of $\theta=90^{\circ}$.

We are planning to make our code available to the public 
so that a wide range of people may use it. 
We believe our code will be very useful for investigating astrophysical and 
data analysis issues of EMRI and for analyzing the data obtained from LISA, 
DECIGO and BBO.

%%%%%%%%%%%%%%%%%%%%%%%%%%%%%%%%%%%%%%%%%%%%%%%%%%%%%%%%%%%%%%%%
\section*{Acknowledgements}
We would like to thank T.~Tanaka for useful discussions and comments. 
We are also grateful to M.~Sasaki and F.~Takahara for their continuous support 
and encouragement.
W.H. was supported by a JSPS Research Fellowship for Young Scientists, 
No.~1756. 
H.T's work was supported in part by grants from JSPS, KAKENHI, Nos. 16540251 and 20540271.

%%%%%%%%%%%%%%%%%%%%%%%%%%%%%%%%%%%%%%%%%%%%%%%%%%%%%%%%%%%%%%%%
%   Appendix
%%%%%%%%%%%%%%%%%%%%%%%%%%%%%%%%%%%%%%%%%%%%%%%%%%%%%%%%%%%%%%%%
\appendix

%%%%%%%%%%%%%%%%%%%%%%%%%%%%%%%%%%%%%%%%%%%%%%%%%%%%%%%%%%%%%%%%
\section{Explicit Expressions for $Z^{\infty}_{\ell m\omega}$ and $Z^{\T{H}}_{\ell m\omega}$}\label{sec:teukolsky-formalism}
%%%%%%%%%%%%%%%%%%%%%%%%%%%%%%%%%%%%%%%%%%%%%%%%%%%%%%%%%%%%%%%%

In this section, we give explicit expressions for the amplitude of the partial waves
$Z^{H/\infty}_{\ell m\omega}$.

Using the Green function of the radial Teukolsky equation,
the solutions of the Teukolsky equation are expressed as
\begin{eqnarray}
R_{\ell m \omega}(r)=
R^{\rm up}_{\ell m\omega}(r){Z}^{\infty}_{\ell m \omega}(r)
+ R^{\rm in}_{\ell m\omega}(r){Z}^{\rm H}_{\ell m \omega}(r),
\end{eqnarray}
where $R^{\rm in/up}_{\ell m\omega}(r)$
satisfy ingoing/outgoing wave conditions at the horizon/infinity, and 
\begin{align}
Z^{\T{H}}_{\ell m\omega} &= \frac{\mu B^{\T{trans}}_{\ell m\omega}}
{2i\omega C^{\T{trans}}_{\ell m\omega}B^{\T{inc}}_{\ell m\omega}}
\int^{\infty}_{-\infty}dt e^{i\omega t-im\phi(t)} \C{I}_{\ell
m\omega}^{\T{H}}[r(t),\theta(t)],\cr
Z^{\infty}_{\ell m\omega} &= \frac{\mu}{2i\omega B^{\T{inc}}_{\ell m\omega}}
\int^{\infty}_{-\infty}dt e^{i\omega t-im\phi(t)} \C{I}_{\ell
m\omega}^{\infty}[r(t),\theta(t)],
\label{eq:Z2}
\end{align}
where 
\begin{align}
 \C{I}^{\T{H}}_{\ell m\omega} =&
\left[
R^{\T{up}}_{\ell m\omega}\left\{
A_{nn0}+A_{\bar{m}n0}+A_{\bar{m}\bar{m}0}
\right\}\right.\cr
&\left.-
\frac{dR^{\T{up}}_{\ell m\omega}}{dr}
\left\{
A_{\bar{m}n1}+A_{\bar{m}\bar{m}1}
\right\}
+
\frac{d^2R^{\T{up}}_{\ell m\omega}}{d^2r}
A_{\bar{m}\bar{m}2}
\right]_{r=r(t),\theta=\theta(t)},\cr
 \C{I}^{\infty}_{\ell m\omega} =&
\left[
R^{\T{in}}_{\ell m\omega}\left\{
A_{nn0}+A_{\bar{m}n0}+A_{\bar{m}\bar{m}0}
\right\}\right.\cr
&\left.-
\frac{dR^{\T{in}}_{\ell m\omega}}{dr}
\left\{
A_{\bar{m}n1}+A_{\bar{m}\bar{m}1}
\right\}
+
\frac{d^2R^{\T{in}}_{\ell m\omega}}{d^2r}
A_{\bar{m}\bar{m}2}
\right]_{r=r(t),\theta=\theta(t)}.
\end{align}
Here 
\begin{eqnarray}
A_{nn0}&=& \frac{-2}{\sqrt{2\pi}\Delta^2}
C_{nn}\rho^{-2}\bar{\rho}^{-1}\mathcal{L}_{1}^{\dagger}
\left\{\rho^{-4}\mathcal{L}_2^{\dagger}(\rho^3 S^{a\omega}_{\ell m})\right\},\cr
A_{\bar{m}n0}&=& \frac{2}{\sqrt{\pi}\Delta}
C_{\bar{m}n}\rho^{-3}\left[(\mathcal{L}^{\dagger}_2 S^{a\omega}_{\ell m})
\left(\frac{iK}{\Delta}+\rho+\bar{\rho}\right)
-a\sin\theta S^{a\omega}_{\ell m}\frac{K}{\Delta}(\bar{\rho}-\rho)\right],\cr
A_{\bar{m}\bar{m}0} &=& -\frac{1}{\sqrt{2\pi}}\rho^{-3}\bar{\rho}
C_{\bar{m}\bar{m}}S^{a\omega}_{\ell m}\left[-i\left(\frac{K}{\Delta}\right)_{,r}
-\frac{K^2}{\Delta^2}+2i\rho\frac{K}{\Delta}\right],\cr
A_{\bar{m}n1}&=& \frac{2}{\sqrt{\pi}\Delta}\rho^{-3}
C_{\bar{m}n}\left[\mathcal{L}^{\dagger}_2S^{a\omega}_{\ell m}
+ia\sin\theta(\bar{\rho}-\rho)S^{a\omega}_{\ell m}\right],\cr
A_{\bar{m}\bar{m}1}&=& -\frac{2}{\sqrt{2\pi}}\rho^{-3}\bar{\rho}
C_{\bar{m}\bar{m}}S^{a\omega}_{\ell m}\left(i\frac{K}{\Delta}+\rho\right),\cr
A_{\bar{m}\bar{m}2}&=&-\frac{1}{\sqrt{2\pi}}\rho^{-3}\bar{\rho}
C_{\bar{m}\bar{m}}S^{a\omega}_{\ell m}, 
\end{eqnarray}
where 
$
\mathcal{L}^{\dagger}_\sigma 
  \equiv \partial_\theta - {m}/{\sin\theta}
  +a\omega \sin\theta +\sigma\cot\theta
$
and
\begin{eqnarray}
 C_{nn} &\equiv & 
\frac{\rho^2\overline{\rho}^2}{4\dot{t}}
\left[\mathcal{E}(r^2+a^2)-a\mathcal{L}+\frac{dr}{d\lambda}\right]^2,\cr
 C_{\bar{m}n} &\equiv & -\frac{\rho^2\overline{\rho}}{2\sqrt{2}\dot{t}}
\left[\mathcal{E}(r^2+a^2)-a\mathcal{L}+\frac{dr}{d\lambda}\right]
\left[i\sin\theta\left(a\mathcal{E}-\frac{\mathcal{L}}{\sin^2\theta}\right)
-\frac{1}{\sin\theta}\frac{d\cos\theta}{d\lambda}\right],\cr
 C_{\bar{m}\bar{m}} &\equiv & 
\frac{\rho^2}{2\dot{t}}\left[i\sin\theta\left(a\mathcal{E}-\frac{\mathcal{L}}{\sin^2\theta}\right)
-\frac{1}{\sin\theta}\frac{d\cos\theta}{d\lambda}\right]^2.
\end{eqnarray}

%%%%%%%%%%%%%%%%%%%%%%%%%%%%%%%%%%%%%%%%%%%%%%%%%%%%%%%%%%%%%%%%
\section{Geodesic Motion in the Kerr Spacetime}\label{sec:geodesic-motion-kerr}
%%%%%%%%%%%%%%%%%%%%%%%%%%%%%%%%%%%%%%%%%%%%%%%%%%%%%%%%%%%%%%%%

In this section, we discuss the solutions of the geodesic equations in detail. 
First, we describe analytical solutions of the $r$ and $\theta$ components
of the geodesic equations, which are expressed as
\begin{eqnarray}
&&\left(\frac{dr}{d\lambda}\right)^2 =
R(r), \label{eq:geodR} \\
&&\left(\frac{d\cos\theta}{d\lambda}\right)^2 =
\Theta(\cos\theta),
\label{eq:geodTheta}
\end{eqnarray}
where 
\begin{align*}
P(r) &=\C{E}(r^2+a^2)-a\C{L}_{z},\cr
R(r) &=[P(r)]^2-\Delta[r^2+(a\C{E}-\C{L}_{z})^2+\C{C}], \cr
\Theta(\cos\theta)&=
\C{C} - (\C{C}+a^2(1-\C{E}^2)+\C{L}_{z}^2)\cos^2\theta
+ a^2(1-\C{E}^2)\cos^4\theta.
\end{align*}

Since $R(r)$ and $\Theta(\cos\theta)$ are fourth-order polynomials, 
there are four zeros of $r$ and $\cos\theta$ for each function, respectively.
A geodesic can be specified if we set two zero points,
$r_{\T{min}}$ and $r_{\T{max}}$, for the radial part and one zero point,
$\cos\theta_{\T{min}}$, for the polar part.
This corresponds to the fact that a one-to-one correspondence exists
between $(r_{\T{min}},r_{\T{max}},\theta_{\T{min}})$ and
$(\C{E},\C{L}_{z},\C{C})$.

It is convenient to introduce the orbital parameters, 
eccentricity $e$, semilatus rectum $p$ and inclination angle 
$\theta_{\T{inc}}$, defined as
\begin{align}
 r_{\T{min}} = \frac{p}{1+{e}},\quad
 r_{\T{max}} = \frac{p}{1-{e}},\quad
\theta_{\T{inc}} + (\T{sgn}\, \C{L}_{z})\, \theta_{\T{min}} = \frac{\pi}{2}.
\label{eq:epthetainc}
\end{align}
The three constants of motion, $(\C{E},\C{L}_{z},\C{C})$, are expressed in terms of these 
orbital parameters $(p,{e},\theta_{\T{inc}})$ \cite{Schmidt:2002, Drasco:2006}.

To solve the differential equations for $r$ and $\cos\theta$,
we rewrite $R(r)$ and $\Theta(\cos\theta)$ as
\begin{align*}
R(r) &=(1 - \C{E}^2)(r_1 - r)(r - r_2)(r - r_3)(r - r_4), \cr
\Theta(\cos\theta)&=\C{L}_{z}^2\epsilon_{0}(z_{-}-\cos^2\theta)(z_{+}-\cos^2\theta),
\end{align*}
where 
\begin{align}
 r_{1} = \frac{p}{1-{e}},\quad
 r_{2} &= \frac{p}{1+{e}},\quad
 r_{3} = \frac{(A+B)+\sqrt{(A+B)^2-4AB}}{2},\quad
 r_{4} = \frac{AB}{r_3},\cr
 A+B &= \frac{2}{1-{\C{E}}^2} - (r_1+r_2),\quad
 AB  = \frac{a^2\C{C}}{(1-{\C{E}}^2)\,r_1r_2},
\end{align}
and $\epsilon_{0}=a^2(1-\C{E}^2)/\C{L}_{z}^2,$
$z_{-}=\cos^2\theta_{\T{min}},$ $z_{+}=\C{C}/(\C{L}_{z}^2\epsilon_0z_{-})$.

Let the solutions of Eqs. (\ref{eq:geodR}) and (\ref{eq:geodTheta}) in terms of 
$r$ or $\theta$ be $\lambda^{(r)}(r)$ and $\lambda^{(\theta)}(\theta)$,
respectively. The functions $\lambda^{(r)}(r)$ and $\lambda^{(\theta)}(\theta)$ are
expressed as 
\begin{align}
  \lambda^{(r)}(r) &=
  \left\{\begin{array}{ll} 
  \lambda^{(r)}_0(r)    \,\,\,\,\, &r:r_2\rightarrow r_1,\\
  2\lambda^{(r)}_0(r_1)-\lambda^{(r)}_0(r)    \,\,\,\,\, &r:r_1\rightarrow r_2,
  \end{array}\right.\cr
  \lambda^{(\theta)}(\theta) &=
  \left\{\begin{array}{ll} 
  \lambda^{(\theta)}_0(\theta)    \,\,\,\,\, &\theta:\frac{\pi}{2}\rightarrow \theta_{\rm min},\\
  2\lambda^{(\theta)}_0(\theta_{\rm min})-\lambda^{(\theta)}_0(\theta)    \,\,\,\,\, &\theta:\theta_{\rm min}\rightarrow \pi-\theta_{\rm min},\\
  4\lambda^{(\theta)}_0(\theta_{\rm min})+\lambda^{(\theta)}_0(\theta)    \,\,\,\,\, &\theta:\pi-\theta_{\rm min}\rightarrow \frac{\pi}{2},
  \end{array}\right.
\label{eq:sol_eom1}
\end{align}
where
\begin{align}
 \lambda^{(r)}_0(r) &= \frac{1}{\sqrt{1-\C{E}^2}}\frac{2}{\sqrt{(r_1-r_3)(r_2-r_4)}}
F\left(\arcsin\sqrt{\frac{r_1-r_3}{r_1-r_2}\frac{r-r_2}{r-r_3}}
,\sqrt{\frac{r_1-r_2}{r_1-r_3}\frac{r_3-r_4}{r_2-r_4}}\right),\cr
 \lambda^{(\theta)}_0(\theta) &= \frac{1}{\C{L}_{z}\sqrt{\epsilon_0 z_{+}}}
 F\left(\arcsin\frac{\cos\theta}{\sqrt{z_{-}}},\sqrt{\frac{z_{-}}{z_{+}}}\right),
\end{align}
and the function $F$ is an elliptic integral of the first kind.
In the following, we describe the elliptic integrals and functions 
using the notation in Ref.~\citen{Recipes}.

The orbital frequencies of the radial and polar motion
with respect to $\lambda$, which are denoted by $\Upsilon_{r}$ and $\Upsilon_{\theta}$, respectively,
are defined in (\ref{frequencies}). 
They are expressed as
\begin{align}
 \Upsilon_{r} =
 \frac{\pi\sqrt{(1-{\C{E}}^2)(r_1-r_3)(r_2-r_4)}}{2K(k_r)},\quad
\Upsilon_{\theta} =
\frac{\pi\C{L}_{z}\sqrt{\epsilon_0 z_{+}}}{2K(k_{\theta})}.
\label{eq:fre_r}
\end{align}
Here $K(k)$ is the complete elliptic integral of the first kind, and 
\begin{align}
 k_{r} = \sqrt{\frac{r_1-r_2}{r_1-r_3}\frac{r_3-r_4}{r_2-r_4}},\quad
k_{\theta} =\sqrt{\frac{z_{-}}{z_{+}}}.
\end{align}

Furthermore, Eq.~\eqref{eq:sol_eom1} can be solved inversely.
\begin{align}
 r(w_r)=\frac{r_3(r_1-r_2)\,\T{sn}^2(\varphi_r(w_r);k_r)-r_2(r_1-r_3)}
 {(r_1-r_2)\,\T{sn}^2(\varphi_r(w_r);k_r)-(r_1-r_3)},\quad
[\cos\theta](w_\theta) = \sqrt{z_{-}}\,\T{sn}(\varphi_\theta(w_\theta);k_\theta).
\label{eq:ana_r}
\end{align}
For convenience, we have introduced the angle variables,
\begin{align}
 w_{r} = \Upsilon_{r}\lambda,\quad
 w_{\theta} = \Upsilon_{\theta}\lambda.
\end{align}
The function $\T{sn}(\varphi,k)$ 
(and $\T{cn}(\varphi,k)$ and  $\T{dn}(\varphi,k)$ below) 
is a Jacobi elliptic function and
\begin{align}
  \varphi_r(w_r)&=
  \left\{\begin{array}{ll} 
  w_r\frac{K(k_r)}{\pi}, \,\,\,\,\, &(0\le w_r \le \pi)\\
  (2\pi-w_r)\frac{K(k_r)}{\pi}, \,\,\,\,\, &(\pi\le w_r \le 2\pi)
  \end{array}\right.\cr
  \varphi_\theta(w_\theta)&=
  \left\{\begin{array}{ll} 
  w_\theta\frac{2K(k_\theta)}{\pi}, \,\,\,\,\, &(0\le w_\theta \le \frac{\pi}{2})\\
  (\pi-w_\theta)\frac{2K(k_\theta)}{\pi}, \,\,\,\,\, &(\frac{\pi}{2}\le w_\theta \le \frac{3\pi}{2})\\
  (w_\theta-2\pi)\frac{2K(k_\theta)}{\pi}. \,\,\,\,\, &(\frac{3\pi}{2}\le w_\theta \le \pi)
  \end{array}\right.
\end{align}
By differentiating $r$ and $\cos\theta$ with respect to $\lambda$, 
we respectively obtain $dr/d\lambda$ and
$d\cos\theta/d\lambda$ analytically, which are expressed as
\begin{align}
 \left[\frac{dr}{d\lambda}\right](w_r)=&
2 {\rm sgn}\left(\frac{d\varphi_r}{dw_r}\right)\cr
&\times\frac{\T{sn}(\varphi_r;k_r)\T{cn}(\varphi_r;k_r)
 \T{dn}(\varphi_r;k_r)(r_2-r_3)(r_1-r_3)(r_1-r_2)}
 {((r_1-r_2)\,\T{sn}^2(\varphi_r;k_r)-(r_1-r_3))^2}
\frac{K(k_r)\Upsilon_r}{\pi}
,\cr
\left[\frac{d\cos\theta}{d\lambda}\right](w_\theta)=&
2 {\rm sgn}\left(\frac{d\varphi_\theta}{dw_\theta}\right)
\sqrt{z_-}\T{cn}(\varphi_\theta;k_\theta)\T{dn}(\varphi_\theta;k_\theta)
\frac{K(k_\theta)\Upsilon_\theta}{\pi},
\label{eq:ana_dr}
\end{align}
where 
\begin{eqnarray*}
{\rm sgn}(x)=
  \left\{\begin{array}{ll} 
  1 \,\,\,\,\, &{\rm for}\,\,\, x > 0,\\
  0 \,\,\,\,\, &{\rm for}\,\,\, x=0,\\
  -1 \,\,\,\,\, &{\rm for}\,\,\, x < 0.
  \end{array}\right.
\end{eqnarray*}

\end{document}